\documentclass[11pt]{article}
\pdfoutput=1
\usepackage{jheppubmod}
\usepackage[punctsep]{collref}
\usepackage{graphicx}
\usepackage{amssymb}
\usepackage{amsmath,amssymb}
\usepackage{slashed}
\usepackage{hyperref}
\usepackage{caption}
\usepackage{xcolor}
\usepackage{dsfont}
\usepackage{verbatim}
\usepackage{subfig}
\usepackage{amsmath,amssymb,amscd,amsfonts,mathtools}
\usepackage{xcolor}
\definecolor{markgreen}{RGB}{230,243,230}
\definecolor{darkolivegreen}{rgb}{0.33, 0.42, 0.18}
\definecolor{darkpastelgreen}{rgb}{0.01, 0.75, 0.24}

\DeclareMathOperator{\arccosh}{arccosh}

\usepackage[toc,page]{appendix}
\usepackage{epsfig}
\usepackage{epstopdf}
\usepackage{latexsym}
\usepackage{graphicx}
\usepackage{booktabs}
\usepackage{bbm}
\usepackage{color}
\usepackage{physics}
\usepackage{tensor}
\usepackage{verbatim}
\usepackage{subfig}
\usepackage{tikz}
\usepackage{ifthen}
\usetikzlibrary{matrix}
\usetikzlibrary{decorations.markings,calc,shapes,decorations.pathmorphing,patterns,decorations.pathreplacing,arrows.meta}
\usetikzlibrary{positioning}
\usepackage{hyperref}
\usetikzlibrary{patterns}
\usetikzlibrary{positioning}

\pdfoutput=1
\makeatletter
\def\@fpheader{\relax}
\makeatother

\newboolean{showcomments}
\setboolean{showcomments}{false}
\newcommand\rem[1]{\ifthenelse{\boolean{showcomments}}{{#1}}{}}
\newcommand{\be}{\begin{equation}}
\newcommand{\ee}{\end{equation}}

\title{\Large Holography and Causality in the Karch-Randall Braneworld}
\preprint{today}
\abstract{It has been argued that the existence of a geodesic shortcut in the Karch-Randall (KR) braneworld rules out the possibility for a low-energy intermediate theory, even when  gravity is turned off.  We study this problem in an explicit example with two symmetrically placed KR branes.   We find that there can be a consistent quantization of bulk matter fields with a holographic intermediate description.  In our model, there is no causality violation. We use this example to study potential implications of the bulk geodesic shortcut. We find that there are possible enhancements of the correlation or entanglement between the two branes that is directly correlated with the extra-dimensional geometry. This could affect the low-energy regime when there are interactions between low and high energy modes. Our model   makes it clear that any possible causality violation can only arise from the UV. Independently of the quantization, an intermediate description should be valid as an EFT.

}

\author{Hao Geng, Lisa Randall}
\affiliation{Gravity, Spacetime, and Particle Physics Initiative, Harvard University, 17 Oxford St., Cambridge, MA, 02138, USA.}
\emailAdd{haogeng@fas.harvard.edu, randall@g.harvard.edu}
\begin{document}
\maketitle
\flushbottom
\pagebreak

\section{Introduction}
The Karch-Randall (KR) braneworld model 
 has several intriguing features, which have made it a useful holographic framework for the study of interesting physical questions and systems, including the black hole information paradox \cite{Almheiri:2019hni,Almheiri:2019psy,Geng:2020qvw,Chen:2020uac,Chen:2020hmv,Grimaldi:2022suv} and boundary conformal field theories \cite{Fujita:2011fp,Takayanagi:2011zk}. The recent progress in understanding  the black hole information paradox in the KR framework relies on the so-called intermediate description, which contains a gravitational system coupled to a nongravitational bath. The gravitational system is the Karch-Randall brane and the nongravitational bath is the asymptotic boundary of the ambient bulk spacetime. This theory has been generalized to include two branes that describe two intermediate-dimensional theories that interact.

However, the consistency of the intermediate description as a local low energy effective theory, even in the absence  of quantum gravity, was challenged by \cite{Omiya:2021olc} based on the existence of a geodesic shortcut in the ambient spacetime that connects the brane and the boundary \cite{Geng:2020np}. Naively, this bulk geodesic shortcut seems to allow a shorter proper time for information to be transferred from the brane to the boundary, in apparent violation of  intermediate description causality, where timelike or spacelike should be determined solely in the four-dimensional theory.\footnote{In this paper, we will implicitly consider the setup with four dimensional Karch-Randall branes in a five-dimensional ambient bulk spacetime. However, the results persist in any dimension.}

In this paper, we address this concern. We argue that such a geodesic shortcut does not interfere with a consistent causal intermediate description for a low-energy effective theory. In the specific case of two symmetrically placed branes, we present a consistent holographic description that captures the transparent boundary condition.  Because the low-energy theory involves only modes with four-dimensional propagation, there is no five-dimensional light-cone singularity in their correlators, which would have been the hallmark of causality violation since it indicates non-commuting spacelike operators in the four-dimensional description.
In our quantization procedure, there would not be causality violation even when heavy modes are included due to the constraints imposed by unitarity. 

The other interesting aspect of constructing an explicit description is that we can see how enhanced correlation (or entanglement) can emerge from the contribution of the heavy modes.
This is a beautiful example of how entanglement can potentially grow in parallel with the emergence of a fifth dimension \cite{VanRaamsdonk:2010pw}. Although a finite number of heavy modes does not yield a nonzero commutator (and hence does not violate causality), as each term in the sum vanishes, we will see that the sum over the contributions from heavy modes can yield an enhanced correlator for light modes coupling to the heavy modes that in essence probes the five-dimensional bulk geometry.

The would-be shortcut could nonetheless potentially lead to causality violation in different quantizations with a less restrictive unitarity constraint, as we will explain.  In such a situation, if light to heavy couplings exist,  the contribution of the infinite sum of heavy modes could potentially allow for the singularity that indicates causality violation. More generally, the UV modes contribute to enhanced correlation or entanglement when they couple to the low-energy theory. The delicate tuning of coefficients required to yield  a singularity would in any case be sensitive to the UV cutoff in any UV completion of the 4d theory as well as the precise implementation of the transparent boundary condition. We do not know if this is ever a concern, but it is of interest to see how it might arise.

We construct a holographic description based on two independent four-dimensional theories corresponding to the theories on the branes and introduce an interaction reflecting the transparent boundary condition.  We work in the $G_N\rightarrow 0$ limit where the causality puzzle that we are addressing resides. Note that strictly speaking in a gravitational universe diffeomorphism invariance would require dressed operators and not strictly local operators. However, the naive tension between the causal structures of the bulk and intermediate descriptions, due to the geodesic shortcut, is at zeroth order in $G_{N}$. Thus, the questions we address refer to the leading O($G_{N}^{0}$) limit. For this reason, we trust the low-energy sector of our analysis. We emphasize that there can be large corrections in the UV, which is why the high-energy result would be sensitive to the cutoff.

We first show how the transparent boundary condition can be implemented in a purely 4d theory. We then introduce a bulk scalar field (or more than one) and treat the KK modes as degrees of freedom in the intermediate theory and introduce consistent boundary conditions according to the transparency for each of the two four-dimensional theories. Note that our intermediate theory is not the usual effective theory obtained by integrating over the extra dimension, which involves only one bulk field with a single set of KK modes that cannot properly reflect the transparent boundary condition.
The  geometry of each brane is AdS$_{4}$ and so it has a dual three-dimensional CFT description, and the transparent boundary condition is a marginal double-trace deformation that couples the two CFTs corresponding to the two branes.  Any five-dimensional bulk field, on the other hand, can be decomposed into four-dimensional modes.  These modes are irreducible representations of the $AdS_4$ isometry group and therefore should describe single-particle states in AdS$_{4}$. These KK modes correspond to primary operators of the three-dimensional CFT. 
 Thus, their propagation from one brane to the other naturally probes the intermediate description causal structure. This propagation is described by two-point correlators of the field which creates such one-particle states. Moreover, these KK modes are dual to primary operators of the three-dimensional CFT.

From a three-dimensional point of view, the transparent boundary condition is a boundary coupling between the conformal field theories associated with the two branes. From the four-dimensional point of view, we find that the light fields on the left brane acts as a (dynamical) source for the field on the right brane and vice versa.  From the five-dimensional point of view, the transparent boundary condition corresponds to the five-dimensional field creating and annihiliating modes on both branes, generating nonvanishing correlation functions for fields on the two branes whose modes are entangled.

In this formulation, it is manifest that bulk locality, i.e. the bulk geodesic, can at best be encoded in  heavy KK modes as each individual mode sees only the metric of the four-dimensional space on which we perform the decomposition and violations of microcausality as determined by the nonzero commutator for spacelike separation can be only a result of UV contributions.  As a result, the existence of the bulk geodesic shortcut does not invalidate the intermediate description as a local effective theory in the low-energy regime. We demonstrate furthermore that even with the inclusion of high-energy degrees of freedom in our quantization, any potential noncausal propagation does not appear because of constraints imposed on the KK modes by unitarity. Since we know the KK modes should ultimately reproduce the 5d theory, the sum of the KK contributions to the bulk two-point function must give a consistent bulk propagator. We will show how our formulation reproduces the singularity in the bulk two-point function corresponding to the 5d light-cone. Such a singularity could only emerge in   the 4d theory when all the modes can experience the transparent boundary condition.

 We  note that others \cite{Neuenfeld:2023svs,Karch:2022rvr,Mori:2023swn} have addressed some of these issues.  When considered from a purely UV string-theory perspective,  the theory is causal by construction.  However, there is clear relationship between the bulk and an intermediate description.  Here we take an approach based on gauge-invariant single-trace operators in an explicit low-energy field theory. Our work is more similar in spirit to \cite{Neuenfeld:2023svs} who clarified the geodesic shortcut from \cite{Omiya:2021olc}. However, they used a cutoff for the effective theory based on where 4d behavior breaks down and argue that any potential causality violation necessarily occurs only above that scale. We do not impose this cutoff. Nonetheless, we show explicitly in our example what the implication of the UV can be and find that whether or not causality is violated can be determined only with better knowledge of the UV structure of the theory or equivalently a consistent quantization that works to all energies. We find the low-energy effective theory preserves the intermediate description causality.

This paper is organized as follows. In Sec.~\ref{sec:puzzle} we  outline the causality puzzle that we are trying to resolve. In Sec.~\ref{sec:review} we study the physics in a toy model of the intermediate description in which we have two AdS$_{d}$ spaces coupled to each other at their asymptotic boundaries. We compute various two-point correlators for a free scalar field in AdS$_{d}$ that obey the transparent boundary condition near the boundary of one AdS$_{d}$ leaking to another AdS$_{d}$. We will use these correlators to compute the corresponding commutators. The resulting commutators are consistent with the causal structure of the two coupled AdS$_{d}$ spacetimes. In Sec.~\ref{sec:Resolution} we provide a holographic dictionary to extract intermediate description low-energy correlators from properly quantized bulk fields in the KR braneworld. The results are consistent with those in Sec.~\ref{sec:review} and are hence manifest resolutions of the puzzle outlined in Sec.~\ref{sec:puzzle}. In Sec.~\ref{sec:heavy KK modes} we study the effects of heavy KK modes. We found that they are important for the AdS$_{d+1}$ bulk locality and the interaction between these heavy KK modes and low-energy KK modes potentially induces violations of the causality of the intermediate description in the low-energy regime. We will see that this type of causality violation is in fact prohibited by unitarity. In Sec.~\ref{sec:other} we discuss the physics in more general situations in the KR braneworld. In Sec.~\ref{sec:conclusion} we conclude this paper with discussion. We propose a potential extension of our analysis to more general situations in the Appendix.

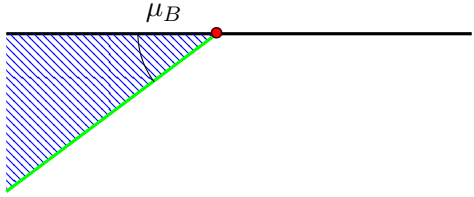
\begin{figure}
\begin{centering}
\begin{tikzpicture}[scale=1.4]
\draw[-,very thick,black!100] (-2,0) to (0,0);
\draw[-,very thick,black!100] (0,0) to (2.42,0);
\draw[pattern=north west lines,pattern color=blue!200,draw=none] (0,0) to (-2,-1.5) to (-2,0) to (0,0);
\draw[-,very thick,color=green!!50] (0,0) to (-2,-1.5);
\draw[-] (-0.75,0) arc (180:217.5:0.75);
\node at (-0.5,0.2) {$\mu_{B}$};
\node at (0,0) {\textcolor{red}{$\bullet$}};
\node at (0,0) {\textcolor{black}{$\circ$}};
\end{tikzpicture}
\caption{The original version of the Karch-Randall braneworld where the AdS$_{d+1}$ is taken to be in the Poincar\'{e} patch and the brane is the green slice which subtends a constant angle $\mu_{L}$ with the asymptotic boundary. The brane tension determines the brane angle. The branes cutoff the bulk shaded regions behind them. The red defect is where the AdS$_{d}$ is glued to a half Minskowski space in the intermediate description. We draw a constant time slice.}
\label{pic:demon}
\end{centering}
\end{figure}

\begin{figure}
\begin{centering}
\begin{tikzpicture}[scale=1.4]
\draw[-,very thick,black!100] (-2,0) to (0,0);
\draw[-,very thick,black!100] (0,0) to (2.42,0);
\draw[pattern=north west lines,pattern color=blue!200,draw=none] (0,0) to (-2,-1.5) to (-2,0) to (0,0);
\draw[pattern=north west lines,pattern color=blue!200,draw=none] (0,0) to (2.42,0) to (2.42,-0.212) to (0,0);
\draw[-,very thick,color=green!!50] (0,0) to (-2,-1.5);
\draw[-,very thick,color=green!!50] (0,0) to (2.42,-0.212);
\draw[-] (-0.75,0) arc (180:217.5:0.75);
\node at (-0.5,0.2) {$\mu_{L}$};
\draw[-] (1.5,0) arc (0:-5.25:1.5);
\node at (1.3,0.2) {$\pi-\mu_{R}$};
\node at (0,0) {\textcolor{red}{$\bullet$}};
\node at (0,0) {\textcolor{black}{$\circ$}};
\node at (-1,-0.75) {\textcolor{black}{$\bullet$}};
\node at (-1,-1) {$P$};
\node at (1,-0.1) {\textcolor{black}{$\bullet$}};
\node at (1,-0.3) {$Q$};
\end{tikzpicture}
\caption{A canonical deformation of the original version of the Karch-Randall braneworld as shown in Fig.~\ref{pic:demon}. Now we have two AdS$_{d}$ branes-- the left and right green slices. They intersect on the red defect on the bulk asymptotic boundary and form a wedge in the bulk. The brane tensions determine the angles $\mu_{L}$ and $\mu_{R}$.}
\label{pic:wedge}
\end{centering}
\end{figure}
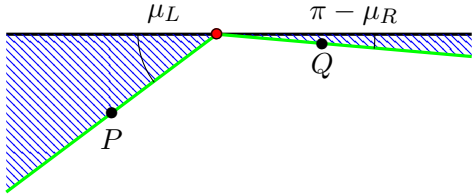

\section{The Causality Puzzle in the Karch-Randall Braneworld}\label{sec:puzzle}

In this section, we outline the causality puzzle by considering the causal structures suggested by the bulk description and by the intermediate description of the KR braneworld. We consider the time it takes to travel between two points along a null geodesic as a kinematical probe of the causal structure in each description and also use this section to set up the notation for later discussions. 

More explicitly, our studies are carried out in a natural deformation of the original KR model. The original setup of the KR model \cite{Karch:2000ct,Karch:2000gx} contains one AdS$_{d}$ brane which is embedded in an ambient AdS$_{d+1}$ spacetime (see Fig.~\ref{pic:demon}). This embedding can be easily understood using the following coordinate in the Poincar\'{e} patch of the ambient AdS$_{d+1}$
\begin{equation}
    ds^{2}=\frac{1}{\sin^{2}\mu}\Big(\frac{du^2+d\vec{x}^2-dt^2}{u^2}+d\mu^2\Big)\,,\label{eq:metricambient}
\end{equation}
where we have set the AdS$_{d+1}$ length scale $l_{\text{AdS}}$ to one, $\vec{x}$ has $(d-2)$ components, $\mu\in(0,\pi)$ and $u\in(0,\infty)$. This coordinate is related to the standard Poinca\'{e} patch coordinate via the following reparametrization \cite{Geng:2020fxl,Karch:2022rvr}
\begin{equation}
    z=u\sin\mu\,,\quad x_{d-1}=u\cos\mu\,.
\end{equation}
In the coordinate Eq.~(\ref{eq:metricambient}), the AdS$_{d}$ brane is embedded along a constant-$\mu$ slice $\mu=\mu_{B}$ with the value of $\mu_{B}$ determined by the tension $T_{B}$ of the brane as
\begin{equation}
    T_{B}=(d-1)\abs{\cos\mu_{B}}\,,
\end{equation}
where we only consider positive tension branes (see Fig.~\ref{pic:demon}) \cite{Geng:2020qvw,Geng:2020fxl,Geng:2021iyq}. With this brane embedded, part of the ambient AdS$_{d+1}$ spacetime will be cut off. This cutoff region is the wedge between the brane and the closest half of the asymptotic boundary of the AdS$_{d+1}$ (the shaded region in Fig.~\ref{pic:demon}).
An interesting deformation of this original KR setup introduces another AdS$_{d}$ brane into the game. In this new setup, the two AdS$_{d}$ branes form a wedge in the ambient AdS$_{d+1}$ spacetime (see Fig.~\ref{pic:wedge}). Similar to the original KR setup, three equivalent descriptions of this deformed setup are generated by applying the AdS/CFT correspondence twice:
\begin{itemize}
    \item \textbf{The Bulk Description:} A quantum gravity theory that lives in this wedge.
    \item \textbf{The Intermediate Description:} Two AdS$_{d}$ quantum gravity theories coupled to each other by transparent boundary conditions through their common asymptotic boundary. Their common asymptotic boundary is a (d-1)-dimensional defect in the bulk (the red dot in Fig.~\ref{pic:wedge}).
    \item \textbf{The Boundary Description:} A (d-1)-dimensional CFT living on the defect.
\end{itemize}
This is also called wedge holography \cite{Geng:2020fxl,Miao:2020oey,Akal:2020wfl}. The existence of the bulk geodesic shortcut is readily seen in this deformed setup. One can take two points on the respective branes ($P$ and $Q$ in Fig.~\ref{pic:wedge}) and tune the values of the two brane tensions in such a way that the two brane angles $\mu_{L}$ and $\mu_{R}$ are close to $\frac{\pi}{2}$ from the below and the above respectively. Thus, the bulk distance between the two points is much shorter than the distance between them in the intermediate description. As a result, if we consider the two points $P$ and $Q$ to be at different times such that they are spacelike separated in the intermediate description, then the small ambient bulk spatial distance between them suggests that they will be timelike separated in the ambient AdS$_{d+1}$. We now do a more careful calculation to show the shortcut persists so long as at least one brane is not exactly along with the bulk boundary.

In this paper, we  consider the bulk geometry to be in the Poincar\'{e} patch of empty AdS$_{d+1}$ for the sake of explicitness. We consider extending our results to generic bulk geometries in Sec.~\ref{sec:other}. Hence, we have two branes with the left brane at $\mu=\mu_{L}$ and the right brane at $\mu=\mu_{R}$  in the following bulk metric
\begin{equation}
    ds^2=\frac{1}{\sin^{2}\mu}\Big(\frac{du^2+d\vec{x}^2-dt^2}{u^2}+d\mu^2\Big)\,,\label{eq:metric1}
\end{equation}
where $\vec{x}$ has $(d-2)$ components, $\mu\in[\mu_{L},\mu_{R}]$ with the branes embedded, $u\in(0,\infty)$ with $u=0$ as the defect and the geometries of constant-$\mu$ slices are in the AdS$_{d}$ Poincar\'{e} patch. Let's take two bulk points $P$ and $Q$ where $P$ is on the left brane and $Q$ is on the right brane. These two bulk points are given by the following coordinate values:
\begin{equation}
    \begin{split}
        P&:\quad \mu=\mu_{L}\,,\quad t=0\,,\quad \vec{x}=0\,,\quad u=u_{P}\,,\\
        Q&:\quad \mu=\mu_{R}\,,\quad t=\Delta t_{\text{bdy}}\,,\quad\vec{x}=0\,,\quad u=u_{Q}\,,
    \end{split}
\end{equation}
where we choose $\Delta t_{\text{bdy}}=u_{Q}+u_{P}$ which is the time it takes for a probe light signal to travel from the point $P$ to the point $Q$ within the geometry of the intermediate description. To compare this time with that through the bulk geodesic, let's consider the following two points
\begin{equation}
      \begin{split}
        P&:\quad \mu=\mu_{L}\,,\quad t=0\,,\quad \vec{x}=0\,,\quad u=u_{P}\,,\\
        Q'&:\quad \mu=\mu_{R}\,,\quad t=\Delta t_{\text{bulk}}\,,\quad\vec{x}=0\,,\quad u=u_{Q}\,,
    \end{split}
\end{equation}
and require them to be connected by a bulk light-like geodesic. This allows us to compute the time $\Delta t_{\text{bulk}}$ for a probe light signal to propagate from the point $P$ to the point $Q'$ which has the same spatial coordinates as the point $Q$.

The calculation can be easily performed using the embedding space formalism in the bulk AdS$_{d+1}$. The embedding space is a $(d+3)$-dimensional Minkowski space with metric
\begin{equation}
    ds^{2}=-dX_{0}^{2}-dX_{d+1}^{2}+dX_{d}^{2}+\sum_{a=1,\cdots,d-1}dX_{a}^{2}\,.
\end{equation}
The embedding space coordinates obey the embedding space constraint
\begin{equation}
    -X_{0}^2-X_{d+1}^{2}+X_{d}^{2}+\sum_{a=1,\cdots,d-1}X_{a}^{2}=-1.
\end{equation}
The embedding space coordinates for $P$ and $Q'$ are
\begin{equation}
    \begin{split}
        P:X_{0}&=\frac{u_{P}^2+1}{2u_{P}\sin\mu_{L}}\,,\\
          X_{a}&=-\cot(\mu_{L}) \omega_{a}\,,\\
          X_{d}&=\frac{u_{P}^2-1}{2u_{P}\sin\mu_{L}}\,,\\
          X_{d+1}&=0\,,
    \end{split}
\end{equation}
and
\begin{equation}
    \begin{split}
        Q':X'_{0}&=\frac{u_{Q}^2-\Delta t_{\text{bulk}}^2+1}{2u_{Q}\sin\mu_{R}}\,,\\
          X'_{a}&=-\cot(\mu_{R}) \omega_{a}\,,\\
          X'_{d}&=\frac{u_{Q}^2-\Delta t_{\text{bulk}}^2-1}{2u_{Q}\sin\mu_{R}}\,,\\
          X'_{d+1}&=\frac{\Delta t_{\text{bulk}}}{u_{Q}\sin\mu_{R}}\,,
    \end{split}
\end{equation}
where $\omega_{a}$ are the polar coordinates for $S^{d-1}$. In the embedding space formalism, it is easy to compute the geodesic distance between the two points $P$ and $Q'$ as
\begin{equation}
\begin{split}
D&=\text{arccosh}(X_{0}X'_{0}+X_{d+1}X'_{d+1}-X_{d}X'_{d}-\sum_{a=1,\cdots,d-1}X_{a}X'_{a})\\
&=\text{arccosh}\Bigg(\frac{2u_{P}^2+2u_{Q}^2-2\Delta t_{\text{bulk}}^2-4u_{Q}u_{P
    }\cos\mu_{L}\cos\mu_{R}}{4u_{Q}u_{P}\sin\mu_{L}\sin\mu_{R}}\Bigg)\,.\label{eq:shortcut}
    \end{split}
\end{equation}
A probe light signal will travel along a light-like geodesic which has $D=0$. Therefore, we can solve for $\Delta t_{\text{bulk}}$ as
\begin{equation}
    \Delta t_{\text{bulk}}=\sqrt{u_{P}^2+u_{Q}^2-2u_{P}u_{Q}\cos(\mu_{L}-\mu_{R})}\leq u_{P}+u_{Q}=\Delta t_{\text{bdy}}\,,\label{eq:timedifference}
\end{equation}
where the equality is saturated if and only if $\mu_{L}-\mu_{R}=\pi$ (i.e. $\mu_{L}=0$ and $\mu_{R}=\pi$) or one of $u_{P}$ and $u_{Q}$ equals zero. 

This result suggests that for two branes at nonzero angles in the bulk, the bulk light signal travels faster than light signals in the intermediate description between any two points with one on each brane. This observation indicates inconsistent causal structures between the intermediate description and the bulk description and so a potential causality violation for the intermediate description \cite{Geng:2020np}. However, geodesics are in fact not physical probes of the causal structure as it does not indicate any dynamics. In this paper, we will show that when we use dynamical probes such as matter field correlators, this apparent tension between the bulk and the intermediate descriptions' causal structures can be avoided for low-energy modes.

\section{ A Consistent Model of the Intermediate Description}\label{sec:review}

In this section, we consider what might  encompass the essential physics of the intermediate description from a purely intermediate dimensional perspective (without regards to the higher-dimensional embedding), in hopes of shedding light on how a holographic dictionary between the bulk and the intermediate descriptions should work. We consider a simple model for the intermediate description of the two-brane setup in Fig.~\ref{pic:wedge}.

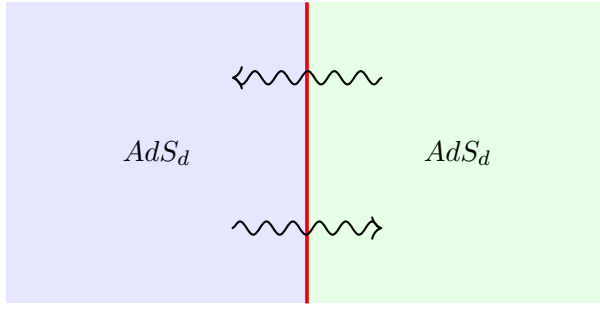
\begin{figure}[h]
    \centering
   \begin{tikzpicture}
      \draw[-,very thick,red](0,-2) to (0,2);
       \draw[fill=green, draw=none, fill opacity = 0.1] (0,-2) to (4,-2) to (4,2) to (0,2);
           \draw[-,very thick,red](0,-2) to (0,2);
       \draw[fill=blue, draw=none, fill opacity = 0.1] (0,-2) to (-4,-2) to (-4,2) to (0,2);
       \node at (-2,0)
       {\textcolor{black}{$AdS_{d}$}};
        \node at (2,0)
       {\textcolor{black}{$AdS_{d}$}};
       \draw [-{Computer Modern Rightarrow[scale=1.25]},thick,decorate,decoration=snake] (-1,-1) -- (1,-1);
       \draw [-{Computer Modern Rightarrow[scale=1.25]},thick,decorate,decoration=snake] (1,1) -- (-1,1);
    \end{tikzpicture}
    \caption{Two AdS$_{d}$ spacetimes coupled to each other near their common asymptotic boundary (the red vertical line). The coupling is achieved by imposing the transparent boundary condition for the matter fields. }\label{pic:interme}
\end{figure}

In this model, we have two AdS$_{d}$ spacetimes with equal-mass free scalar fields 
$\phi_{1}(x)$ and $\phi_{2}(x)$ living respectively on each of them and these two AdS$_{d}$ spacetimes are coupled to each other by a transparent boundary condition between the two fields $\phi_{1}(x)$ and $\phi_{2}(x)$ near their common asymptotic boundary (see Fig.~\ref{pic:interme}). 
Our discussion is self-contained with the notations mostly following Ref. \cite{Geng:2023qwm} so we refer interested readers to Ref. \cite{Geng:2023qwm} for more details and discussions and \cite{Porrati:2001db,Porrati:2001gx,Porrati:2003sa,Kiritsis:2006hy} for relevant earlier works. 

\subsection{The Transparent Boundary Condition}\label{sec:transpabc}
Let's first study how the transparent boundary condition is imposed. We will take the two AdS$_{d}$ length scales to one. The two AdS$_{d}$ spacetimes are described by the metrics
\begin{equation}
    ds_{i}^{2}=\frac{dz_{i}^{2}+\eta_{ab}dx_{i}^{a}dx_{i}^{b}}{z_{i}^{2}}\,,\quad \text{where } i=1,2\,,
\end{equation}
where $z_{i}\in (0,\infty)$ and they share the common asymptotic boundary $z_{1}=z_{2}=0$. We will use $x$ to collectively denote the coordinates $(z,x^{a})$. The total action for the two scalar fields is given by
\begin{equation}
\begin{split}
    S_{\text{tot}}=&-\int d^{d}x_{1}\sqrt{-g}\Big[-\phi_{1}(x_{1})g^{\mu\nu}\partial_{\mu}\partial_{\nu}\phi_{1}(x_{1})+m^{2}\phi_{1}^{2}(x_{1})\Big]\\&-\int d^{d}x_{2}\sqrt{-g}\Big[-\phi_{2}(x_{2})g^{\mu\nu}\partial_{\mu}\partial_{\nu}\phi_{2}(x_{2})+m^{2}\phi_{2}^{2}(x_{2})\Big]\,,
    \end{split}
\end{equation}
where we note that we didn't add any boundary terms to the action because we want to impose the transparent boundary condition. The precise boundary condition for the fields $\phi_{1}(x)$ and $\phi_{2}(x)$ can be obtained by setting the on-shell variation of the total action to zero. This implies that
\begin{equation}
    \delta S_{\text{tot}}^{\text{on-shell}}=\int d^{d-1}x \Big[\frac{1}{z_{1}^{d-2}}\Big(\phi_{1}\partial_{z_{1}}\delta\phi_{1}-\delta\phi_{1}\partial_{z_{1}}\phi_{1}\Big)+\frac{1}{z_{2}^{d-2}}\Big(\phi_{2}\partial_{z_{2}}\delta\phi_{2}-\delta\phi_{2}\partial_{z_{2}}\phi_{2}\Big)\Big]\Big|_{z_{1},z_{2}\rightarrow0}=0\,.
\end{equation}

Within the usual AdS/CFT correspondence, we have as we extrapolate to the boundary
\begin{equation} 
\phi(z,x^a)\sim \frac{J(x^a)} {2 \Delta-d+1} z^{d-1-\Delta}+\hat{O}(x^a) z^{\Delta}\,,\quad \text{as } z\rightarrow0\,,\label{eq:quantization}
\end{equation}
where $\hat{O}(x^{a})$ is the dual CFT operator at the boundary and the function $J(x^{a})$ is the background source. In the \textit{standard quantization} scheme one takes $\Delta=\Delta_{+}$ and in the \textit{alternative} quantization scheme we have $\Delta=\Delta_{-}$, where $\Delta_{\pm}=\frac{d-1}{2}\pm\sqrt{m^{2}+\frac{(d-1)^{2}}{4}}$. In out case, the near asymptotic boundary expansion of a generic on-shell configuration of the scalar fields is
\begin{equation}
\begin{split}
    \phi_{1}(x_{1})&\sim z_{1}^{d-1-\Delta_{+}} \Big[\beta_{1}(x_{1}^{a})+\mathcal{O}(z_{1}^{2})\Big]+z_{1}^{\Delta_{+}} \Big[\alpha_{1}(x_{1}^{a})+\mathcal{O}(z_{1}^{2})\Big]\,,\\\phi_{2}(x_{2})&\sim z_{2}^{d-1-\Delta_{-}} \Big[\beta_{2}(x_{2}^{a})+\mathcal{O}(z_{2}^{2})\Big]+z_{2}^{\Delta_{-}} \Big[\alpha_{2}(x_{2}^{a})+\mathcal{O}(z_{2}^{2})\Big]\,,\label{eq:asymp}
    \end{split}
\end{equation}
where we consider the case $\frac{d+1}{2}>\Delta_{+}>\frac{d-3}{2}$ for which both the $\alpha_{i}$ modes and $\beta_{i}$ modes are normalizable in the Klein-Gordon norm \cite{Geng:2023ynk}. Notice that this means that for this quantization we are using both the standard and the alternative quantizations. As we shall see, the sources are promoted to dynamical fields. Observe that $\beta_1$, the source term on the first brane, is dynamical in this setup because of the transparent boundary condition, which involves the product of left and right CFT operators, allowing the field on one brane to source the field on the other brane. Near the asymptotic boundary, we can identify $x_{1}^{a}=x_{2}^{a}=x^{a}$ and thus we have
\begin{equation}
\begin{split}
    \delta S_{\text{tot}}^{\text{on-shell}}&=\int d^{d-1}x \Big[\frac{1}{z_{1}^{d-2}}\Big(\phi_{1}\partial_{z_{1}}\delta\phi_{1}-\delta\phi_{1}\partial_{z_{1}}\phi_{1}\Big)+\frac{1}{z_{2}^{d-2}}\Big(\phi_{2}\partial_{z_{2}}\delta\phi_{2}-\delta\phi_{2}\partial_{z_{2}}\phi_{2}\Big)\Big]\Big|_{z_{1},z_{2}\rightarrow0}\,,\\&=\int d^{d-1}x (\Delta_{-}-\Delta_{+})\Big(\alpha_{1}\delta\beta_{1}-\delta\alpha_{1}\beta_{1}-\alpha_{2}\delta\beta_{2}+\delta\alpha_{2}\beta_{2}\Big)\,,\\&=0\,,
    \end{split}
\end{equation}
which is solved by
\begin{equation}
    \beta_{1}=c\alpha_{2}\,,\quad\beta_{2}=-c\alpha_{1}\,,\label{eq:resulttrans}
\end{equation}
where we will see later that $c=\frac{g}{2\Delta_{+}-d+1}$ with the coefficients $g$ parametrizes the strength of the coupling, as in Eq.~(\ref{eq:CFTaction}) below. Moreover, both the $\alpha$-modes and the $\beta$-modes must be dynamical, i.e. nonzero. Otherwise the energy-momentum flux near the asymptotic boundary from one AdS$_{d}$ to the other AdS$_{d}$ would be zero \cite{Geng:2023ynk}. 

The result Eq.~(\ref{eq:resulttrans}) is  consistent with the dual CFT description \cite{Geng:2023ynk}. In the dual CFT description, we have two CFT$_{d-1}$'s that are respectively dual to the gravitational theories on two AdS$_{d}$'s and they are coupled together by a marginal double-trace deformation. The total action in this CFT description is
\begin{equation}
S_{\text{tot}}=S_{\text{CFT$_{1}$}}+S_{\text{CFT$_{2}$}}+g\int d^{d-1}x O_{1}(x)O_{2}(x)\,,\label{eq:CFTaction}
\end{equation}
where $g$ denotes the coupling constants, $\phi_{1}(x)$ is a single-trace scalar operator in the CFT$_{1}$ and $\phi_{2}(x)$ is a single-trace scalar operator in the CFT$_{2}$. The conformal weight of the operator $O_{1}(x)$ is $\Delta_{+}$ and that of $O_{2}(x)$ is $\Delta_{-}$. Therefore, we have $\Delta_{+}+\Delta_{-}=d-1$ so the coupling in Eq.~(\ref{eq:CFTaction}) is a marginal deformation which is consistent with the fact that in the KR braneworld the intermediate description is dual to a boundary CFT, i.e. the boundary description. The AdS/CFT correspondence states that the operator $\phi_{1}(x)$ duals to the scalar field $\phi_{1}(x)$ and the operator $O_{2}(x)$ duals to the scalar field $\phi_{2}(x)$.

Furthermore, the coupling in Eq.~(\ref{eq:CFTaction}) suggests that from the CFT$_{1}$ point of view $gO_{2}(x)$ is the source of the operator $O_{1}(x)$ and from the CFT$_{2}$ point of view $gO_{1}(x)$  is the source of the operator $O_{2}(x)$. Thus, the two CFT$_{d-1}$'s are mutually sourcing each other and one controls the dynamics of the source for the other. In our case, the source $J$ as defined in Equ.~(\ref{eq:quantization}) gets promoted to be dynamical which we denote using the mode $\beta$.

Translating the above observation back to the AdS language using the AdS/CFT correspondence we have 

\begin{equation}
\begin{split}
    \beta_{1}(x^{a})&=\frac{g}{2\Delta_{+}-d+1}O_{2}(x^{a})\,,\quad \alpha_{1}(x^{a})=O_{1}(x^{a})\,,\\ \beta_{2}(x^{a})&=\frac{g}{2\Delta_{-}-d+1}O_{1}(x^{a})\,,\quad\alpha_{2}(x^{a})=O_{2}(x^{a})\,,\label{eq:transp}
    \end{split}
\end{equation}
which accounts for the proportionality constant we chose in  Eq.~(\ref{eq:resulttrans}).

\subsection{Canonical Quantization and Correlators}

Having discussed the fields at the boundary, we now extend our analysis to the full four-dimensional space.

Since both the $\alpha$-modes and the $\beta$-modes are dynamical, in the canonical quantization of the fields $\phi_{i}(x)$ we have two sets of creation and annihilation operators. This is as opposed to the reflective boundary conditions in AdS/CFT where either $\alpha$-modes or $\beta$-modes are dynamical but not both and we have only a single set of creation and annihilation operators.

We assign creation and annihilation operators for each on-shell mode such that the canonical quantization condition is obeyed. In the Poincar\'{e} patch of AdS$_{d}$, the on-shell modes for a scalar field with mass squared $m^{2}$ are 
\begin{equation}
    \phi_{\omega,\vec{k}}^{\pm}(x)=z^{\frac{d-1}{2}} J_{\pm\sqrt{\frac{(d-1)^2}{4}+m^{2}}}(z\sqrt{\omega^{2}-\vec{k}^{2}})e^{i\omega t-\i\vec{k}\cdot \vec{x}}\,,\label{eq:adsdmodes}
\end{equation}
where $J_{\nu}(x)$ is the Bessel function of the first kind, we have included both positive and negative $\nu$, and we have $\omega^{2}-\vec{k}^{2}\geq0$. We have the near asymptotic boundary expansions for these modes
\begin{equation}
    \phi^{\pm}_{\omega,\vec{k}}(x)\sim z^{\Delta_{\pm}} \frac{1}{2^{\pm\sqrt{\frac{(d-1)^2}{4}+m^{2}}}\Gamma[1\pm\sqrt{\frac{(d-1)^2}{4}+m^{2}}]} e^{i\omega t-i\vec{k}\cdot \vec{x}}+\mathcal{O}(z^{2})\,,\quad\text{as } z\rightarrow0\,.\label{eq:asympmodes}
\end{equation}
Therefore, the canonical quantization of the two fields $\phi_{i}(x_{i})$ is given by
\begin{equation}
    \begin{split}
      \hat{\phi}_{1}(x)&=\int_{\omega^2-\vec{k}^2\geq0} \frac{d\vec{k}}{(2\pi)^{d-1}}\frac{d\omega}{2\pi}\Big(\phi^{+}_{\omega,\vec{k}}(x)\hat{a}^{\dagger}_{\omega,\vec{k}}+\frac{g\phi^{-}_{\omega,\vec{k}}(x)}{2\Delta_{+}-d+1}\hat{b}^{\dagger}_{\omega,\vec{k}}+\phi^{+*}_{\omega,\vec{k}}(x)\hat{a}_{\omega,\vec{k}}+\frac{g\phi^{-*}_{\omega,\vec{k}}(x)}{2\Delta_{+}-d+1}\hat{b}_{\omega,\vec{k}}\Big)\,,\\
\hat{\phi}_{2}(x)&=\int_{\omega^2-\vec{k}^2\geq0} \frac{d\vec{k}}{(2\pi)^{d-1}}\frac{d\omega}{2\pi}\Big(\frac{g\phi^{+}_{\omega,\vec{k}}(x)}{2\Delta_{-}-d+1}\hat{a}^{\dagger}_{\omega,\vec{k}}+\phi^{-}_{\omega,\vec{k}}(x)\hat{b}^{\dagger}_{\omega,\vec{k}}+\frac{g\phi^{+*}_{\omega,\vec{k}}(x)}{2\Delta_{-}-d+1}\hat{a}_{\omega,\vec{k}}+\phi^{-*}_{\omega,\vec{k}}(x)\hat{b}_{\omega,\vec{k}}\Big)\,,\label{eq:canonicalquantization}
    \end{split}
\end{equation}
where  $\omega$ is constrained to obey $\omega\geq 0$ in the above integrals and the commutator between the creation and annihilation operators are determined by the canonical quantization condition with the two sets of creation and annihilation operators mutually commuting with each other. It's easy to see that Eq.~(\ref{eq:canonicalquantization}) is consistent with Eq.~(\ref{eq:asymp}).\footnote{More precisely, using Eq.~(\ref{eq:asympmodes}), we have 
\begin{equation}
\begin{split}
        \alpha_{1}&=\frac{g}{2\Delta_{+}-d+1}\frac{1}{2^{-\sqrt{\frac{(d-1)^2}{4}+m^{2}}}\Gamma[1-\sqrt{\frac{(d-1)^2}{4}+m^{2}}]}\int_{\omega^2-\vec{k}^2\geq0} \frac{d\vec{k}}{(2\pi)^{d-1}}\frac{d\omega}{2\pi}\Big( e^{i\omega t-i\vec{k}\cdot \vec{x}}\hat{b}^{\dagger}_{\omega,\vec{k}}+ e^{-i\omega t+i\vec{k}\cdot \vec{x}}\hat{b}_{\omega,\vec{k}}\Big)\,,\\
        \beta_{1}&=\frac{1}{2^{\sqrt{\frac{(d-1)^2}{4}+m^{2}}}\Gamma[1+\sqrt{\frac{(d-1)^2}{4}+m^{2}}]}\int_{\omega^2-\vec{k}^2\geq0} \frac{d\vec{k}}{(2\pi)^{d-1}}\frac{d\omega}{2\pi}\Big( e^{i\omega t-i\vec{k}\cdot \vec{x}}\hat{a}^{\dagger}_{\omega,\vec{k}}+ e^{-i\omega t+i\vec{k}\cdot \vec{x}}\hat{a}_{\omega,\vec{k}}\Big)\,.
\end{split}
\end{equation}
} As a result, we can compute various two-point correlators. For example, the two-point correlator for operator insertions within the first AdS$_{d}$ is
\begin{equation}
    \begin{split}
        \langle \phi_{1}(x_{1})\phi_{1}(y_{1})\rangle=\int\frac{d\vec{k}}{(2\pi)^{d-1}}\frac{d\omega}{2\pi} \frac{d\vec{k}'}{(2\pi)^{d-1}}\frac{d\omega'}{\sqrt{2\omega'}}&\Big[\phi^{+*}_{\omega,\vec{k}}(x_{1})\phi^{+}_{\omega',\vec{k}'}(y_{1})\bra{0}\hat{a}_{\omega,\vec{k}}\hat{a}^{\dagger}_{\omega',\vec{k}'}\ket{0}\\&+g^{2}\frac{\phi^{-*}_{\omega,\vec{k}}(x_{1})\phi^{-}_{\omega',\vec{k}'}(y_{1})}{(2\Delta_{+}-d+1)^{2}}\bra{0}\hat{b}_{\omega,\vec{k}}\hat{b}^{\dagger}_{\omega',\vec{k}'}\ket{0}\Big]\,,\label{eq:11inte}
    \end{split}
\end{equation}
for which the canonical commutation relation gives us 
\begin{equation}
     \langle \phi_{1}(x_{1})\phi_{1}(y_{1})\rangle=\frac{1}{(2\Delta_{+}-d+1)^{2}+g^{2}}\Big[(2\Delta_{+}-d+1)^{2}G^{+}(x_{1},y_{1})+g^{2}G^{-}(x_{1},y_{1})\Big]\,,\label{eq:11}
\end{equation}
where we have the standard bulk AdS$_{d}$ Green's functions
\begin{equation}
    G^{\pm}(x_{1},y_{1})=\frac{2^{\frac{d}{2}-1}\Gamma[\Delta_{\pm}]}{\pi^{\frac{d-1}{2}}\Gamma[\Delta_{\pm}-\frac{d-3}{2}]}\frac{_{2}F_{1}[\frac{\Delta_{\pm}}{2},\frac{\Delta_{\pm}+1}{2},\Delta_{\pm}-\frac{d-3}{2},\frac{1}{Z_{11}^{2}}]}{(2Z_{11})^{\Delta_{\pm}}}\,,\label{eq:G}
\end{equation}
with the invariant distance 
\begin{equation}
Z_{11}=\frac{z_{1}^{2}+w_{1}^{2}-(t_{1}-\tau_{1})^{2}+i\epsilon \text{ sgn}(t_{1}-\tau_{1})+(\vec{x}_{1}-\vec{y}_{1})\cdot (\vec{x}_{1}-\vec{y}_{1})}{2z_{1}w_{1}}\,, \label{eq:Z11}   
\end{equation} 
where we used $x_{1}=(z_{1},t_{1},\vec{x}_{1})$ and $y_{1}=(w_{1},\tau_{1},\vec{y}_{1})$ and the $i\epsilon$ prescription regularizes the $\omega$ integrals in Eq.~(\ref{eq:11inte}) with $\epsilon\rightarrow0^{+}$. Similarly, we also have the two-point correlator for operator insertions within the second AdS$_{d}$ as
\begin{equation}
    \begin{split}
        \langle \phi_{2}(x_{2})\phi_{2}(y_{2})\rangle=\int\frac{d\vec{k}}{(2\pi)^{d-1}}\frac{d\omega}{2\pi} \frac{d\vec{k}'}{(2\pi)^{d-1}}\frac{d\omega'}{\sqrt{2\omega'}}&\Big[g^{2}\frac{\phi^{+*}_{\omega,\vec{k}}(x_{1})\phi^{+}_{\omega',\vec{k}'}(y_{1})}{(2\Delta_{-}-d+1)^{2}}\bra{0}\hat{a}_{\omega,\vec{k}}\hat{a}^{\dagger}_{\omega',\vec{k}'}\ket{0}\\&+\phi^{-*}_{\omega,\vec{k}}(x_{1})\phi^{-}_{\omega',\vec{k}'}(y_{1})\bra{0}\hat{b}_{\omega,\vec{k}}\hat{b}^{\dagger}_{\omega',\vec{k}'}\ket{0}\Big]\,,
    \end{split}
\end{equation}
for which the canonical commutation relation gives us 
\begin{equation}
     \langle \phi_{2}(x_{2})\phi_{2}(y_{2})\rangle=\frac{1}{(2\Delta_{-}-d+1)^{2}+g^{2}}\Big[g^{2}G^{+}(x_{2},y_{2})+(2\Delta_{-}-d+1)^{2}G^{-}(x_{2},y_{2})\Big]\,,\label{eq:22}
\end{equation}
where $x_{2}=(z_{1},t_{1},\vec{x}_{2})$ and $y_{2}=(w_{2},\tau_{2},\vec{y}_{2})$ and we should use the invariant distance 
\begin{equation}
Z_{22}=\frac{z_{2}^{2}+w_{2}^{2}-(t_{2}-\tau_{2})^{2}+i\epsilon \text{ sgn}(t_{2}-\tau_{2})+(\vec{x}_{2}-\vec{y}_{2})\cdot (\vec{x}_{2}-\vec{y}_{2})}{2z_{2}w_{2}}\,,  \label{eq:Z22}     
\end{equation} 
with $\epsilon\rightarrow0^{+}$. Moreover, we also have the two-point correlator for operators inserted in different AdS$_{d}$'s
\begin{equation}
\begin{split}
     \langle \phi_{1}(x_{1})\phi_{2}(x_{2})\rangle=g\int\frac{d\vec{k}}{(2\pi)^{d-1}}\frac{d\omega}{2\pi} \frac{d\vec{k}'}{(2\pi)^{d-1}}\frac{d\omega'}{\sqrt{2\omega'}}&\Big[\frac{\phi^{+*}_{\omega,\vec{k}}(x_{1})\phi^{+}_{\omega',\vec{k}'}(x_{2})}{2\Delta_{-}-d+1}\bra{0}\hat{a}_{\omega,\vec{k}}\hat{a}^{\dagger}_{\omega',\vec{k}'}\ket{0}\\&+\frac{\phi^{-*}_{\omega,\vec{k}}(x_{1})\phi^{-}_{\omega',\vec{k}'}(x_{2})}{2\Delta_{+}-d+1}\bra{0}\hat{b}_{\omega,\vec{k}}\hat{b}^{\dagger}_{\omega',\vec{k}'}\ket{0}\Big]\,,
     \end{split}
\end{equation}
for which the canonical commutation relation and the fact that $\Delta_{+}+\Delta_{-}=d-1$ give us 
\begin{equation}
     \langle \phi_{1}(x_{1})\phi_{2}(x_{2})\rangle=\frac{g(2\Delta_{-}-d+1)}{(2\Delta_{-}-d+1)^{2}+g^{2}}\Big[G^{+}(x_{1},x_{2})-G^{-}(x_{1},x_{2})\Big]\,,\label{eq:12}
\end{equation}
for which we should use the invariant distance \begin{equation}
Z_{12}=\frac{z_{1}^{2}+z_{2}^{2}-(t_{1}-t_{2})^{2}+i\epsilon \text{ sgn}(t_{1}-t_{2})+(\vec{x}_{1}-\vec{x}_{2})\cdot (\vec{x}_{1}-\vec{x}_{2})}{2z_{1}z_{2}}\,,\label{eq:Z12}       
\end{equation} 
with $\epsilon\rightarrow0^{+}$. 

As a consistency check, we can see that the $g\rightarrow0$ limit of the results Eq.~(\ref{eq:11}), Eq.~(\ref{eq:22}) and Eq.~(\ref{eq:12}) reduce to the correlators for two decoupled AdS$_{d}$'s with the scalar field in the first AdS$_{d}$ under the standard quantization and the scalar field in the second AdS$_{d}$ under the alternative quantization.

\subsection{Analytic Structures of the Correlators and Implications for Causality}\label{sec:intermedcausal}
From the correlators Eq.~(\ref{eq:11}), Eq.~(\ref{eq:12}) and Eq.~(\ref{eq:22}), we can compute the expectation values of the commutators. These commutators encode the causal structure of the theory as they should be zero when the two points are spacelike separated and potentially nonzero if the two points are timelike separated for a causal theory. Such a transition for the commutators from being zero to being potentially nonzero, when the two points are deformed from being spacelike separated to being timelike separated, is due to the existence of a branchcut of the two-point correlators when one fixes the coordinates of one point and treats the coordinates of the other point as variables. For a causal theory, such branchcuts should start  at a singularity corresponding to the place when the two points are null separated. Hence, to understand the causal structure of the theory we will locate the branchcuts of the two-point correlators and thus the singularity of the two-point correlators. 

\begin{figure}[h]
    \centering
   \begin{tikzpicture}
      \draw[->,very thick,black] (0,-3) to (0,3);
      \draw[-,very thick,black] (-2,0) to (0,0);
      \draw[->,very thick, black] (2,0) to (4,0);
      \node at (0,0) {\textcolor{black}{$\bullet$}};
    \node at (2,0) {\textcolor{black}{$\bullet$}};
    \node at (-2,-0) {\textcolor{black}{$\bullet$}};
      \draw[-,very thick] 
       decorate[decoration={zigzag,pre=lineto,pre length=5pt,post=lineto,post length=5pt}] {(0,0) to (2,0)};
       \draw[-,very thick] 
       decorate[decoration={zigzag,pre=lineto,pre length=5pt,post=lineto,post length=5pt}] {(-4,0) to (-2,0)};
       \node at (-0.7,3) {\text{Im(Z)}};
       \node at (4,-0.5) {\text{Re(Z)}};
       \node at (2,0.3) {\text{$1$}};
       \node at (-2.1,0.3) {\text{$-1$}};
    \end{tikzpicture}
    \caption{The branchcuts of the correlators $G^{\pm}$ as a function of the invariant distance Z. The two branchcuts are along the real axis of $Z$ starting respectively at the singular points $Z=\pm1$ of $G^{\pm}$ and terminate at $Z=0$ and $Z=-\infty$ which is another singular point of $G^{\pm}$.}\label{pic:branchcut}
\end{figure}
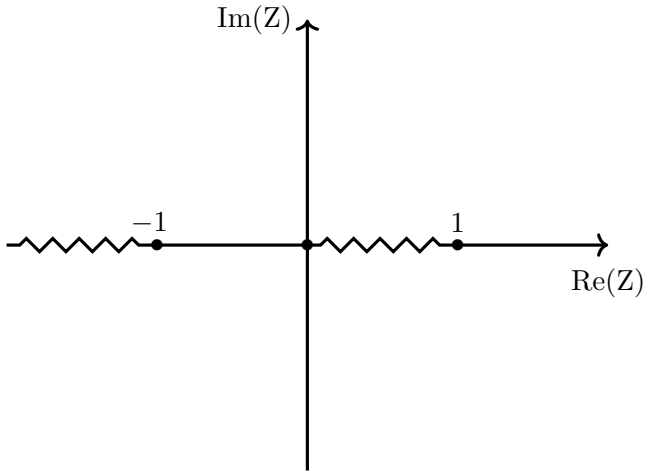

Using the basic properties of the hypergeometric functions, one can see that the two correlators $G^{\pm}$  have branchcuts only starting at the place where the invariant distance takes the value $\pm 1$ which are also the singularities of the correlators $G^{\pm}$. One of these branchcuts ends on another singular point of the correlators $G^{\pm}$ for which the invariant distance is zero and the other branchcut ends at infinity. We will take the branchcuts for the two correlators $G^{\pm}$ as functions of the invariant distance to be along the real axis with one goes from $Z=1$ to $Z=0$ and the other goes from $Z=-1$ to $Z=-\infty$ (see Fig.~\ref{pic:branchcut}). 

The two singularities at $Z=\pm1$ have clear physical meanings. For two points within the same AdS$_{d}$, the $Z=1$ singularity means this two points are null separated as the geodesic distance between them is $D=\arccosh(Z)=\arccosh(1)=0$ and the $Z=-1$ singularity means that one point and the mirror-point with respect to the asymptotic boundary, i.e. $z\rightarrow-z$, of the other point are null separated as their geodesic distance $D=\arccosh(-Z)=\arccosh(1)=0$.\footnote{This latter singularity at $Z=-1$ in fact suggests the existence of the reflection of the scalar wave by the asymptotic boundary whose physics will be discussed in Sec.~\ref{sec:tranmrefl}.} For two points in different AdS$_{d}$'s, the singularity $Z=1$ suggests an unphysical light-cone since as we will see it would generate nonzero commutators for two spacelike separated points, and the singularity at $Z=-1$ means that the two points in respective AdS$_{d}$'s are null separated as their geodesic distance is $D=\arccosh(-Z)=\arccosh(1)=0$.

Now we are ready to compute various commutators. The commutators are potentially nonzero if and only if the shortest line connecting invariant distance $Z$ and its complex conjugation in the complex $Z$ plane crosses a branchcut. The detailed results are as follows.
\begin{itemize}
    \item \textbf{The commutator within the first AdS$_{d}$}: 
    
    We first compute $\langle[\phi_{1}(x_{1}),\phi_{1}(y_{1})]\rangle$ to outline some technical details and latter we will use these results to obtain other commutators. Since the invariant distance $Z_{11}$ in Eq.~(\ref{eq:Z11}) with an $i\epsilon$ prescription depends on the ordering of the inserted operators, we have
    \begin{equation}
          \langle \phi_{1}(y_{1})\phi_{1}(x_{1})\rangle=\langle\phi_{1}(x_{1})\phi_{1}(y_{1})\rangle^{*}\,.
    \end{equation}
Therefore the commutator can be written as
\begin{equation}
\begin{split}
    \langle[\phi_{1}(x_{1}),\phi_{1}(y_{1})]\rangle=\langle\phi_{1}(x_{1})\phi_{1}(y_{1})\rangle-\langle\phi_{1}(x_{1})\phi_{1}(y_{1})\rangle^{*}\,.\label{eq:commutator11}
    \end{split}
\end{equation}
From Eq.~(\ref{eq:11}), Eq.~(\ref{eq:G}) and Eq.~(\ref{eq:Z11}), we can see that the commutator Eq.~(\ref{eq:commutator11}) is potentially nonzero if and only if the invariant distance $Z_{11}$ is vertically below or above a branchcut, i.e.
\begin{equation}
    0<\text{Re}Z_{11}<1\,,\quad \text{or } \text{Re}Z_{11}<-1\,\,,
\end{equation}
which is equivalent to
\begin{equation}
\begin{split}
    -1&<\frac{(z_{1}-w_{1})^{2}-(t_{1}-\tau_{1})^{2}+(\vec{x}_{1}-\vec{y}_{1})\cdot (\vec{x}_{1}-\vec{y}_{1})}{2z_{1}w_{1}}<0\,,\\&\text{or }\frac{(z_{1}+w_{1})^{2}-(t_{1}-\tau_{1})^{2}+(\vec{x}_{1}-\vec{y}_{1})\cdot (\vec{x}_{1}-\vec{y}_{1})}{2z_{1}w_{1}}<0\,,
    \end{split}
\end{equation}
and can be written equally as
\begin{equation}
\begin{split}
    -1&<\frac{(z_{1}-w_{1})^{2}-(t_{1}-\tau_{1})^{2}+(\vec{x}_{1}-\vec{y}_{1})\cdot (\vec{x}_{1}-\vec{y}_{1})}{2z_{1}w_{1}}<0\,,\\&\text{or }\frac{(z_{1}-w_{1})^{2}-(t_{1}-\tau_{1})^{2}+(\vec{x}_{1}-\vec{y}_{1})\cdot (\vec{x}_{1}-\vec{y}_{1})}{2z_{1}w_{1}}<-2\,,\label{eq:11result}
    \end{split}
\end{equation}
The geodesic distance between the two points is
\begin{equation}
    D(x_{1},y_{1})=\arccosh(\text{Re}Z_{11})\,.
\end{equation}
which implies that the two points $x_{1}$ and $y_{1}$ are timelike separated, i.e. $D(x_{1},y_{1})$ is imaginary, if and only if
\begin{equation}
    \text{Re}Z_{11}<1\,\Longleftrightarrow \frac{(z_{1}-w_{1})^{2}-(t_{1}-\tau_{1})^{2}+(\vec{x}_{1}-\vec{y}_{1})\cdot (\vec{x}_{1}-\vec{y}_{1})}{2z_{1}w_{1}}<0\,.\label{eq:11timelike}
\end{equation}
As a result, from Eq.~(\ref{eq:11result}) and Eq.~(\ref{eq:11timelike}), we conclude that the commutator Eq.~(\ref{eq:commutator11}) is potentially nonzero if and only if the two points $x_{1}$ and $y_{1}$ are timelike separated in the first AdS$_{d}$ spacetime. This result is consistent with causality.

    \item \textbf{The commutator within the second AdS$_{d}$}: 
    
    The commutator $\langle[\phi_{2}(x_{2},\phi_{2}(y_{2}))]\rangle$ within the second AdS$_{d}$ can be computed along the same lines as the commutator within the first AdS$_{d}$. The result is that the commutator is potentially nonzero if and only if the two points $x_{2}$ and $y_{2}$ are timelike separated in the second AdS$_{d}$ spacetime. This result is equally consistent with causality.

    \item \textbf{The commutator across the two AdS$_{d}$'s}:

    As a final test of causality, we have to compute the commutator $\langle[\phi_{1}(x_{1}),\phi_{2}(x_{2})]\rangle$ with $x_{1}$ in the first AdS$_{d}$ and $x_{2}$ in the second AdS$_{d}$. Naively, the branchcut structure in Fig.~\ref{pic:branchcut} suggests that this commutator is potentially nonzero if and only if
    \begin{equation}
        0<\text{Re}Z_{12}<1\,,\quad \text{or } \text{Re}Z_{12}<-1\,\,,
\end{equation}
which is equivalent to
\begin{equation}
\begin{split}
    -1&<\frac{(z_{1}-z_{2})^{2}-(t_{1}-t_{2})^{2}+(\vec{x}_{1}-\vec{x}_{2})\cdot (\vec{x}_{1}-\vec{x}_{2})}{2z_{1}z_{2}}<0\,,\\&\text{or }\frac{(z_{1}+z_{2})^{2}-(t_{1}-t_{2})^{2}+(\vec{x}_{1}-\vec{x}_{2})\cdot (\vec{x}_{1}-\vec{x}_{2})}{2z_{1}z_{2}}<0\,.
    \end{split}
\end{equation}
Nevertheless, in this case the two points $x_{1}$ and $x_{2}$ are timelike separated if and only if
\begin{equation}
    \frac{(z_{1}+z_{2})^{2}-(t_{1}-t_{2})^{2}+(\vec{x}_{1}-\vec{x}_{2})\cdot (\vec{x}_{1}-\vec{x}_{2})}{2z_{1}z_{2}}<0\,,\label{eq:12causalZ}
\end{equation}
which is exactly the branchcut from for $Z_{12}$ from $-1$ to $-\infty$. Hence, causality requires that the branchcut for $Z_{12}$ from $1$ to $0$ should disppear in the two-point correlator $\langle\phi_{1}(x_{1})\phi_{2}(x_{2})\rangle$ in Eq.~(\ref{eq:12}). This can be easily seen to be the case as follows. The two functions $G^{+}$ and $G^{-}$ as functions of the invariant distance $Z_{12}$ are solutions of a certain second order linear differential equation. For the branchcut starting at $Z=1$ to disappear in the linear combination Eq.~(\ref{eq:12}), it is enough to check that there is no singularity at $Z=1$ for Eq.~(\ref{eq:12}). Since it is a solution of a second order linear differential equation, it is enough to check whether the first two singular terms of $G^{+}$ and $G^{-}$ as $Z_{12}\rightarrow1$ cancel in the linear combination Eq.~(\ref{eq:12}). This can be efficiently done using Mathematica and is indeed the case. Therefore, we can conclude that the two-point function $\langle\phi_{1}(x_{1})\phi_{2}(x_{2})\rangle$ only has one branchcut for $Z_{12}$ from $-1$ to $-\infty$.

As a result, the commutator
\begin{equation}
    \langle[\phi_{1}(x_{1}),\phi_{2}(x_{2})]\rangle=\langle\phi_{1}(x_{1})\phi_{2}(x_{2})\rangle-\langle\phi_{1}(x_{1})\phi_{2}(x_{2})\rangle^{*}\,,
\end{equation}
is potentially nonzero if and only if $Z_{12}<-1$ which is equivalent to Eq.~(\ref{eq:12causalZ}), i.e. if and only if $x_{1}$ and $x_{2}$ are timelike separated. This result is again consistent with causality.
\end{itemize}

In summary, we found that the correlators Eq.~(\ref{eq:11}), Eq.~(\ref{eq:22}) and Eq.~(\ref{eq:12}) are all consistent with causality.

\subsection{Reflection and Transmission Coefficients}\label{sec:tranmrefl}

It is interesting to notice from the analysis in Sec.~(\ref{sec:intermedcausal}) that the branchcut for $Z$ from $-1$ to $-\infty$ didn't play an important role in the causal structure for the commutators within each AdS$_{d}$. This is because, for two points $x_{1}=(z_{1},t_{1},\vec{x}_{1})$ and $y_{1}=(w_{1},\tau_{1},\vec{y}_{1})$ within the same AdS$_{d}$, the singularity of the correlator at $Z_{11}=-1$ physically means that the point $x_{1}$ and the mirror-point of $y_{1}$, i.e. $Ry_{1}=(-w_{1},\tau_{1},\vec{y}_{1})$, are null separated and when $Z_{11}<-1$ the two points $x_{1}$ and $Ry_{1}$ are timelike separated which is always satisfied as long as the two points $x_{1}$ and $y_{1}$ are timelike separated. In other words, the lightcone determined by the two points $x_{1}$ and $Ry_{1}$ is always within the lightcone determined by $x_{1}$ and $y_{1}$. This latter lightcone $\text{Re}Z<1$ is controlled by the singularity $Z=1$ so the singularity $Z=-1$ didn't play any essential role in the discussion of the causality within each AdS$_{d}$. Nevertheless, this doesn't mean that the singularity at $Z=-1$ for the within AdS$_{d}$ correlators has no physical implications.

For the within AdS$_{d}$ correlators, the singularity at $Z=-1$ suggests that the scalar wave will be reflected by the asymptotic boundary so the asymptotic boundary is not fully transparent unless the singularity at $Z=-1$, i.e. the branchcut from $Z=-1$ to $-\infty$, disappears from the within AdS$_{d}$ correlators. Hence one can study this reflection by comparing the expansion of these correlators near $Z=1$ and $Z=-1$. We will focus on $\langle\phi_{1}(x_{1})\phi_{1}(y_{1})\rangle$ and the analysis for $\langle\phi_{2}(x_{2})\phi_{2}(y_{2})\rangle$ will follow exactly the same line. We have the following form of the expansions
\begin{equation}
    \begin{split}
\langle\phi_{1}(x_{1})\phi_{1}(y_{1})\rangle&=\sum_{n=0}^{^{\infty}}(a_{n}^{+}+a_{n}^{-})\frac{1}{(Z_{11}-1)^{\frac{d-2}{2}-n}}\,,\quad \text{as }Z_{11}\rightarrow 1\,,\label{eq:Z11near1}
\end{split}
\end{equation}
We note that if $\frac{d}{2}$ is an integer the expansion should involve logarithmic terms. In this paper, for the sake of convenience, we will consider general $\frac{d}{2}\in \mathbb{R}$ and we don't treat integer $\frac{d}{2}$ cases separately as all results persist to these special cases. The expansion in Equ.~(\ref{eq:Z11near1}) can be done using MATHEMATICA and we found that
\begin{equation}
    \frac{a_{n}^{+}}{a_{n}^{-}}=\frac{(2\Delta_{+}-d+1)^{2}}{g^{2}}\,,\quad\forall n\in\mathbb{N}\,.
\end{equation}
We also have
\begin{equation}
\begin{split}
\langle\phi_{1}(x_{1})\phi_{1}(y_{1})\rangle&=\sum_{n=0}^{^{\infty}}((-1)^{-\frac{d}{2}-\Delta_{+}+1+n}a_{n}^{+}+(-1)^{\Delta_{+}-\frac{3d}{2}+n}a_{n}^{-})\frac{1}{(1+Z_{11})^{\frac{d-2}{2}-n}}\,,\quad\text{as } Z_{11}\rightarrow -1\,,
    \end{split}
\end{equation}

We define a reflection coefficient
\begin{equation}
    R=\abs{\frac{(-1)^{-\frac{d}{2}-\Delta_{+}+1+n}a_{n}^{+}+(-1)^{\Delta_{+}-\frac{3d}{2}+n}a_{n}^{-}}{a_{n}^{+}+a_{n}^{-}}}^{2}=\abs{\frac{(2\Delta_{+}-d+1)^{2}+(-1)^{2\Delta_{+}-d+1}g^{2}}{(2\Delta_{+}-d+1)^{2}+g^{2}}}^{2}\leq1\,,\label{eq:reflection}
\end{equation}
which is independent of the order of the expansion $n$. This reflection coefficient is zero for the special case $m^{2}=-\frac{(d-1)^{2}-1}{4}$ and $g=1$ which corresponds to the fully transparent boundary condition for a conformally coupled scalar field. In fact, this is an important consistency check of the definition Eq.~(\ref{eq:reflection}) as a reflection coefficient. The reason is that the AdS$_{d}$ geometry is Weyl equivalent to the a half-Minkowski space so the conformally coupled scalar field living in AdS$_{d}$ is equivalent to a massless scalar field living on a half-Minkowski space for which a fully transparent boundary condition necessarily exists.

Similarly, one can define the transparent coefficient by comparing the near $Z=-1$ expansion of the across AdS$_{d}$'s correlator $\langle\phi_{1}(x_{1})\phi_{2}(x_{2})\rangle$ with the near $Z=1$ expansion of the correlator $\langle\phi_{1}(x_{1})\phi_{1}(y_{1})\rangle$ in Eq.~(\ref{eq:Z11near1}). The former expansion is given by
\begin{equation}
    \langle\phi_{1}(x_{1})\phi_{2}(x_{2})\rangle=\frac{g}{2\Delta_{-}-d+1}\sum_{n=0}^{^{\infty}}((-1)^{-\frac{d}{2}-\Delta_{+}+1+n}a_{n}^{+}+(-1)^{\Delta_{+}-\frac{3d}{2}+n}b_{n}^{-})\frac{1}{(1+Z_{12})^{\frac{d-2}{2}-n}}\,,\quad \text{as }Z_{12}\rightarrow -1\,,
\end{equation}
for which we have from MATHEMATICA
\begin{equation}
    \frac{b_{n}^{-}}{a_{n}^{+}}=-1\,.
\end{equation}
Therefore, we can define the transmission coefficient
\begin{equation}
\begin{split}
    T&=\abs{\frac{g}{2\Delta_{-}-d+1}\frac{(-1)^{-\frac{d}{2}-\Delta_{+}+1+n}a_{n}^{+}+(-1)^{\Delta_{+}-\frac{3d}{2}+n}b_{n}^{-}}{a_{n}^{+}+a_{n}^{-}}}^{2}\,,\\&=\abs{g\frac{(2\Delta_{+}-d+1)}{(2\Delta_{+}-d+1)^{2}+g^{2}}\Big(1-(-1)^{2\Delta_{+}-d+1}\Big)}^{2}\leq1\,,\label{eq:transmission}
    \end{split}
\end{equation}
which is independent of the the order of the expansion $n$ and we used the AM-GM inequality in the last step. 

As a consistency check, we have $T=1$ for the case $m^{2}=-\frac{d^{2}-1}{4}$ and $g=1$ which is exactly the same as the case when $R=0$ and this is the case when the scalar field is conformally coupled to the background metric. Since AdS$_{d}$ is Weyl equivalent to a half Minkowski space, our setup with two glued AdS$_{d}$'s is Weyl equivalent to a Minkowski space. Thus, the conformally coupled scalar field is equivalently living on a Minkowski space with a trivial interface separating the two half Minkowski spaces which is consistent with the results $T=1$ and $R=0$. Moreover, we notice that from Eq.~(\ref{eq:reflection}) and Eq.~(\ref{eq:transmission}) we have $R+T=1$ which lends further credence to our interpretation.

\section{A Causal Holographic Intermediate Description}\label{sec:Resolution}
The previous section presented a procedure to quantize two AdS theories interacting on a common boundary. In the $d-1$-dimensional description, we have two interacting CFTs. In the $d$-dimensional description, there are modes of equal mass whose corresponding CFT operators interact on a common boundary so that the left and right fields source each other.  This yields the boundary conditions for the fields that create or annihilate fields on either AdS.
%, with the linear combination of the two noninteracting fields (or more precisely noninteracting associated CFTs) corresponding to the precise transparent boundary condition. 

In this section, we consider bulk fields whose KK modes interact precisely as the modes did in the previous section. These KK modes are the single particle states in the intermediate description.  Unlike the other two holographic theories, we do not present an interaction between the KK modes on the two branes, or treat one 5d field as sourcing another 5d field, both of which could exist only on the boundary and would presumably require counterterms. Instead, we treat the transparent boundary condition as implemented for the bulk fields by KK modes that create or annihilate fields on either brane. If we take this to be true for the KK modes, it is also true for the bulk fields to which they correspond. Said differently, the transparent boundary condition from the bulk description should be interpreted as entanglement of the modes on the two branes. This is necessary as there is no way to impose a boundary condition on bulk fields that would allow modes to transfer energy from one brane to another without having the bulk  fields create modes on both branes. We will make this more explicit in what follows. 

When we implement this procedure, we take at most one or two bulk fields. Although sufficient to understand the implications of the transparent boundary condition on the 4d fields, it is presumably not adequate to completely describe any particular CFT. Nonetheless it suffices to address the question of whether there is a well-defined theory and if it is causal.

In this section, we take the model of  Sec.~\ref{sec:review} to be  the model of the intermediate description of the full doubly holographic theory. We ask how the physics in the intermediate description is encoded in the bulk description.
To answer this, we propose a holographic dictionary that allows us to extract the various propagators in the transparent sector of the intermediate description from the bulk.

We again model the transparent sector of the intermediate description by weakly coupled scalar fields and see how the various correlators can be extracted from a properly quantized bulk field. This gives a consistent holographic dictionary in the Karch-Randall braneworld for correlators and the resulting intermediate description correlators are  the same as those in Sec.~\ref{sec:review}. In this section, we focus on the symmetric case $\mu_{L}=\pi-\mu_{R}$. With an explicit quantization,  the  resolution of the causality puzzle in Sec.~\ref{sec:puzzle} becomes apparent.

\subsection{The Holographic Dictionary}
In the standard AdS/CFT duality \cite{Maldacena:1997re,Witten:1998qj,Gubser:1998bc}, fields in  AdS are the Kaluza-Klein (KK) modes of the fields in the full 10d supergravity theories reduced on a compact internal manifold. In the prototypical example of Type IIB supergravity on AdS$_{5}\times S^{5}$ \cite{Maldacena:1997re,Aharony:1999ti,Gunaydin:1984fk,Kim:1985ez}, the compact internal manifold is $S^{5}$. We take the scalar fields in the intermediate description of the KR theory to be KK modes of the fields in the bulk description. That is, we take the KK modes to be the holographically dual single particle states corresponding to  CFT primary operators. Such a bottom-up prescription is  motivated in part by top-down constructions of the KR model where $d=4$ and the KR brane captures the region in the bulk whose geometry is AdS$_{4}\times M^{6}$ where $M^{6}$ is a compact internal manifold \cite{Bachas:2017rch,Bachas:2018zmb,Uhlemann:2021nhu}. In that case, the AdS$_{4}$ fields are KK modes of the 10d fields compactified on $M^{6}$.

Our approach will be a bottom-up study of an effective field theory model of holography in which we take the bulk field to be a free scalar field whose KK modes are a tower of free scalar fields in AdS$_{d}$. As in the case of the prototypical example of AdS/CFT, only the low-lying KK modes are reliable operators in the AdS$_{d}$ and they are light fields in the AdS$_{d}$. These AdS$_{d}$ light fields will have boundary conditions that model the transparent sector of the intermediate description. Similar to the standard AdS/CFT studies, this requires a proper quantization of the bulk scalar field. Here we determine whether such a minimal model could capture the structure of the intermediate description and whether it satisfies the intermediate description causal structure.

\subsubsection{Quantization of the Bulk Field}\label{sec:quantizing bulk field}
Let's consider a free massive scalar field $\chi(x,\mu)$ in the AdS$_{d+1}$ bulk with two branes. We will use the metric in Eq.~(\ref{eq:metric1}). The scalar field obeys the equation of motion
\begin{equation}
    \Box_{\text{AdS$_{d+1}$}}\chi-m^{2}\chi=0\,,\label{eq:eom1}
\end{equation}
where $\Box_{\text{AdS$_{d+1}$}}$ is the bulk AdS$_{d+1}$ d'Alembertian operator and $x=(u,t,\vec{x})$ denotes the AdS$_{d}$ coordinates.  The equation of motion Eq.~(\ref{eq:eom}) can be written as
\begin{equation}
  \partial_{\mu}^{2}\chi-(d-1)\cot(\mu)\partial_{\mu}\chi+\Box_{\text{AdS$_d$}}\chi-\frac{1}{\sin^{2}\mu}m^{2}\chi=0\,,
\end{equation}
where $\Box_{\text{AdS$_{d}$}}$ is the AdS$_{d}$ d'Alembertian operator in the Poincar\'{e} patch. To obtain the proper two-point function of $\chi(x,\mu)$ and extract the intermediate description correlators from it, the bulk field $\chi(x,\mu)$ has to be properly quantized. This requires using an appropriate basis for the modes of the field $\chi(x,\mu)$. For our purpose of the holographic study, we decompose the field $\chi(x,\mu)$ into KK modes
\begin{equation}
\chi_{n}^{\pm}(x,\mu)=\psi_{n}(\mu)\chi_{n}(x)=\psi_{n}(\mu)f_{k,\omega,n}(u)e^{i\omega t-i\vec{k}\cdot \vec{x}}\,,
\end{equation}
where we have $\omega^2>\vec{k}^2$ and 
\begin{equation}
f_{k,\omega,n}(u)=u^{\frac{d-1}{2}}J_{\pm\sqrt{\frac{(d-1)^2}{4}+m_{n}^2}}(u\sqrt{\omega^2-\vec{k}^2})\,,
\end{equation}
are the AdS$_{d}$ modes Eq.~(\ref{eq:adsdmodes}) for a free scalar field with $m_{n}$ the corresponding the KK mass.

The basis is specified by the choice of the boundary conditions of the KK wavefunction $\psi_{n}(\mu)$ at the two branes. Our choice is determined by  two considerations. First, as we discussed in Sec.~\ref{sec:transpabc}, in the intermediate description the light matter fields on the respective AdS$_{d}$'s are mutually sourcing each other such that we have a transparent boundary condition between the two AdS$_{d}$'s \cite{Geng:2023ynk}. Second, in the braneworld holography both matter fields and their sources in the dual theory living on the brane are dynamical \cite{Geng:2023qwm}. Moreover, these matter fields are dual to Dirichlet modes of the bulk fields near the brane and their sources are dual to Neumann modes of the same bulk fields near the same brane. These two considerations suggest that for the bulk field we should have KK modes that satisfy different boundary conditions near the two different branes. More specifically, the matter fields on the left AdS$_{d}$ should correspond to the KK modes whose wavefunctions satisfy the Dirichlet (D) boundary condition near the left brane and the Neumann (N) boundary condition on the right brane. Therefore, they can be thought as matter fields on the left AdS$_{d}$ and as sources on the right AdS$_{d}$. We will denote the KK wavefunction of these modes by $\psi_{n}^{DN}(\mu)$. Similarly, we should also have KK modes with wavefunctions $\psi_{n}^{ND}(\mu)$ which can be thought of as sources on the left AdS$_{d}$ and as matter fields on the right AdS$_{d}$. Thus, for each KK mass we should have modes associated with two linearly independent KK wavefunctions. These wavefunctions satisfy the following boundary conditions near the branes
\begin{equation}
\begin{split}
\psi_{n}^{DN}(\mu_{L})&=0\,,\quad\psi_{n}^{DN\prime}(\mu_{R})=0\,,\\\psi_{n}^{ND\prime}(\mu_{L})&=0\,,\quad\psi_{n}^{ND}(\mu_{R})=0\,,\label{eq:bc}
\end{split}
\end{equation}
and their wave equation is
\begin{equation}
\psi_{n}''(\mu)-(d-1)\frac{\cos\mu}{\sin\mu}\psi_{n}'(\mu)+m_{n}^2\psi_{n}(\mu)-\frac{m^2}{\sin^{2}\mu}\psi_{n}(\mu)=0\,.\label{eq:eom}
\end{equation}

For this quantization to work,  the modes associated to different KK wavefunctions that appear in the decomposition of a single field must be degenerate. For general brane angles, this degeneracy is not manifest--a point we return to later on. However,  when $\mu_{L}=\pi-\mu_{R}$, the modes are degenerate. The reason is that for any solution $\psi_{n,1}(\mu)$ of Eq.~(\ref{eq:eom}) $\psi_{n,2}=\psi_{n,1}(\pi-\mu)$ is equally a solution. Moreover, if one takes $\psi_{n,1}(\mu)$ to obey the boundary conditions of $\psi_{n}^{DN}(\mu)$ then $\psi_{n,2}(\mu)$ automatically obeys the boundary condition for $\psi_{n}^{ND}(\mu)$. Therefore, in the case we are considering such a degeneracy is guaranteed by a symmetry.

Now let's discuss a global relation obeyed by the two wavefunctions $\psi_{n}^{DN}(\mu)$ and $\psi_{n}^{ND}(\mu)$. For the sake of convenience of the discussion, we define
\begin{equation}
\psi_{n}(\mu)=\sin^{\frac{d-1}{2}}\mu \Psi_{n}(\mu)\,,\label{eq:redef}
\end{equation}
with which the KK wave equation Eq.~(\ref{eq:eom}) transforms to
\begin{equation}
\begin{split}
\Psi_{n}''(\mu)-\frac{d-1}{2}\Psi_{n}(\mu)&-\frac{d^2-1}{4}\frac{\cos^{2}\mu}{\sin^{2}\mu}\Psi_{n}(\mu)+m_{n}^2\Psi_{n}(\mu)-\frac{m^2}{\sin^{2}\mu}\Psi_{n}(\mu)=0\,.\label{eq:waveeq}
\end{split}
\end{equation}
For the simplicity of the notation, we denote 
\begin{equation}
y_{n,1}=\Psi_{n}^{DN}(\mu)\,,\quad y_{n,2}=\Psi_{n}^{ND}(\mu)\,.   \label{eq:ys}
\end{equation}
Let's consider consider the Wronskian associated to the two solutions of Eq.~(\ref{eq:waveeq}) in Eq.~(\ref{eq:ys})
\begin{equation}
W_{n}(\mu)=y_{n,1}(\mu)y_{n,2}^{\prime}(\mu)-y_{n,1}^{\prime}(\mu)y_{n,2}(\mu)\,.
\end{equation}
Using Eq.~(\ref{eq:waveeq}) we can derive $W_{n}'(\mu)=0$ so we see that $W_{n}(\mu)$ is a constant independent of $\mu$. As a result, we have
\begin{equation}
W_{n}(\mu_{L})=W_{n}(\mu_{R})\,.\label{eq:W}
\end{equation}
The boundary conditions in Eq.~(\ref{eq:bc}) reduce Eq.~(\ref{eq:W}) to
\begin{equation}
    -y_{n,1}'(\mu_{L})y_{n,2}(\mu_{L})=y_{n,1}(\mu_{R})y_{n,2}^{\prime}(\mu_{R})\,.\label{eq:wronskianresult}
\end{equation}
Furthermore, the wave equation Eq.~(\ref{eq:waveeq}) tells us that the wavefunctions $\Psi_{n}^{DN}(\mu)$ and $\Psi_{n}^{ND}(\mu)$ are normalized as
\begin{equation}
    \int_{\mu_{L}}^{\mu_{R}} d\mu \Psi_{n}^{DN}(\mu)\Psi_{m}^{DN}(\mu)=\delta_{nm}\,,\quad\int_{\mu_{L}}^{\mu_{R}} d\mu \Psi_{n}^{ND}(\mu)\Psi_{m}^{ND}(\mu)=\delta_{nm}\,.
\end{equation}

Now we are ready to provide the basis for the modes of the bulk scalar field as
\begin{equation}
\begin{split}
\chi^{a}_{n,\omega,\vec{k}}(x,\mu)=\sin^{\frac{d-1}{2}}\mu\Psi_{n}^{ND}(\mu) u^{\frac{d-1}{2}}J_{\nu_n}(u\sqrt{\omega^2-\vec{k}^2})e^{-i\omega t+i\vec{k}\dot \vec{x}}\,,\\\chi^{b}_{n,\omega,\vec{k}}(x,\mu)=\sin^{\frac{d-1}{2}}\mu\Psi_{n}^{DN}(\mu) u^{\frac{d-1}{2}}J_{-\nu_n}(u\sqrt{\omega^2-\vec{k}^2})e^{-i\omega t+i\vec{k}\dot \vec{x}}\,,\label{eq:convert}
\end{split}
\end{equation}
where $\nu_n=\sqrt{\frac{(d-1)^2}{4}+m_{n}^2}$. Equipped with these modes, the canonically quantized bulk scalar field is given by

\begin{equation}
\begin{split}
\hat{\chi}(x,\mu)=\int\frac{d\vec{k}d\omega}{(2\pi)^{d-1}\sqrt{2\omega}}\sum_{n}\Big[\hat{a}_{n,\omega,\vec{k}}\chi_{n,\omega,\vec{k}}^{a}(x,\mu)+\hat{b}_{n,\omega,\vec{k}}\chi^{b}_{n,\omega,\vec{k}}(x,\mu)+\hat{a}^{\dagger}_{n,\omega,\vec{k}}\chi_{n,\omega,\vec{k}}^{a*}(x,\mu)+\hat{b}^{\dagger}_{n,\omega,\vec{k}}\chi^{b*}_{n,\omega,\vec{k}}(x,\mu)\Big]\,,\label{eq:bulkfield}
\end{split}
\end{equation}
where $\{\hat{a}^{\dagger}_{n,\omega,\vec{k}},\hat{a}_{n,\omega,\vec{k}}\}$ and $\{\hat{b}^{\dagger}_{n,\omega,\vec{k}},\hat{b}_{n,\omega,\vec{k}}\}$ are two independent sets of creation and annihilation operators. Their commutation relations are determined by the bulk canonical quantization condition. A caveat is that  we are only interested in light KK modes in this section and the right hand side of Equ.~(\ref{eq:bulkfield}) should be understood as only for the light KK modes.

The bulk two-point function can now be obtained as
\begin{equation}
    \begin{split}
        \frac{\langle\hat{\chi}(x_{1},\mu_{1})\hat{\chi}(x_{2},\mu_{2})\rangle}{\sin^{\frac{d-1}{2}}\mu_{1}\sin^{\frac{d-1}{2}}\mu_{2}}=&\sum_{n}\Big[\Psi_{n}^{ND}(\mu_{1})\Psi_{n}^{ND}(\mu_{2})G_{d,n}^{+}(x_{1},x_{2})+\Psi_{n}^{DN}(\mu_{1})\Psi_{n}^{DN}(\mu_{2})G_{d,n}^{-}(x_{1},x_{2})\Big]\,,\label{eq:dic}
    \end{split}
\end{equation}
where $G_{d,n}^{\pm}(x_{1},x_{2})$ are the AdS$_{d}$ Green's functions Eq.~(\ref{eq:G}) for scalar field with mass square $m_{n}^2$ with the same $i\epsilon$ prescription as Eq.~(\ref{eq:Z11}). More explicitly, we have $G_{d,n}^{\pm}(x_{1},x_{2})$ given by Eq.~(\ref{eq:G}) with $\Delta_{\pm}=\frac{d-1}{2}\pm\sqrt{\frac{(d-1)^2}{4}+m_{n}^2}$ and 
\begin{equation}
    Z=\frac{u_{1}^{2}+u_{2}^{2}-(t_{1}-t_{2})^{2}+i\epsilon \text{ sgn}(t_{1}-t_{2})+(\vec{x}_{1}-\vec{x}_{2})\cdot (\vec{x}_{1}-\vec{x}_{2})} {2u_{1}u_{2}}\,,
\end{equation} 
in the coordinate patch Eq.~(\ref{eq:metric1}) with $\epsilon\rightarrow0^{+}$. 

We note that our quantization Eq.~(\ref{eq:bulkfield}) is motivated by top-down constructions of the deformed KR model \cite{Uhlemann:2021nhu} where $d=4$ and the two branes model two AdS$_{4}\times M^{6}$ regimes of the bulk. Hence, the $ND$ modes are modeling the bulk modes localized in the right AdS$_{4}\times M^{6}$ region and the $DN$ modes are modeling the bulk modes localized in the left AdS$_{4}\times M^{6}$ region.

\subsubsection{Extracting the Intermediate Description Correlators from the Bulk}
With a quantized bulk field, we can extract the various intermediate description correlators from the bulk description to complete a holographic dictionary for KR. 

We first determine the intermediate description operators from the bulk field $\chi(x,\mu)$, which can be extracted from the bulk field $\chi(x,\mu)$ by extrapolation to the branes and proper projection to the corresponding KK modes. Since the intermediate description has a transparent boundary condition, the operators in the intermediate description generally have both the $a$ modes and $b$ modes. A natural dictionary is
\begin{equation}
    \hat{\Phi}_{L,n}(x)=\frac{\hat{\chi}(x,\mu_{L})+\partial_{\mu}\hat{\chi}(x,\mu_{L})}{\sin^{\frac{d-1}{2}}\mu_{L}}\Big|_{n} \,,\quad\hat{\Phi}_{R,n}(x)=\frac{\hat{\chi}(x,\mu_{R})+\partial_{\mu}\hat{\chi}(x,\mu_{R})}{\sin^{\frac{d-1}{2}}\mu_{R}}\Big|_{n} \,,\label{eq:interoperator}
\end{equation}
where $\hat{\Phi}_{L,n}(x)$ and $\hat{\Phi}_{R,n}(x)$, the KK fields, are respectively light operators on the left AdS$_{d}$ and the right AdS$_{d}$ and $\Big|_{n}$ means we take only the $n$-th KK mode. Now we are ready to compute various correlators in the intermediate description. We first compute the  AdS$_{d}$ correlators 
\begin{equation}
    \begin{split}
        \langle\hat{\Phi}_{L,n}(x_{1})\hat{\Phi}_{L,m}(y_{1})\rangle=\delta_{mn}\Big[\Psi_{n}^{ND}(\mu_{L})^2G_{d,n}^{+}(x_{1},y_{1})+\Psi_{n}^{DN\prime}(\mu_{L})^{2}G_{d,n}^{-}(x_{1},y_{1})\Big]\,,\label{eq:targetLL}
    \end{split}
\end{equation}
\begin{equation}
    \langle\hat{\Phi}_{R,n}(x_{2})\hat{\Phi}_{R,m}(y_{2})\rangle=\delta_{mn}\Big[\Psi_{n}^{ND\prime}(\mu_{R})^2G_{d,n}^{+}(x_{2},y_{2})+\Psi_{n}^{DN}(\mu_{R})^{2}G_{d,n}^{-}(x_{2},y_{2})\Big]\,,\label{eq:targetRR}
\end{equation}
where we have used the boundary condition Eq.~(\ref{eq:bc}) and bulk correlator Eq.~(\ref{eq:dic}). The across AdS$_{d}$ correlator can be similarly computed as
\begin{equation}
    \langle\hat{\Phi}_{L,n}(x_{1})\hat{\Phi}_{R,m}(x_{2})\rangle=\delta_{mn}\Big[\Psi_{n}^{ND}(\mu_{L})\Psi_{n}^{ND\prime}(\mu_{R})G_{d,n}^{+}(x_{1},x_{2})+\Psi_{n}^{DN\prime}(\mu_{L})\Psi_{n}^{DN}(\mu_{R})G_{d,n}^{-}(x_{1},x_{2})\Big]\,,\label{eq:target}
\end{equation}
which can be simplified using Eq.~(\ref{eq:wronskianresult}) and the fact that $\Psi_{n}^{ND}(\mu_{R})=\Psi_{n}^{DN}(\pi-\mu_{R})=\Psi_{n}^{DN}(\mu_{L})$ as
\begin{equation}
    \langle\hat{\Phi}_{L,n}(x_{1})\hat{\Phi}_{R,m}(x_{2})\rangle=\delta_{mn}\Psi_{n}^{ND}(\mu_{L})\Psi_{n}^{ND\prime}(\mu_{R})\Big[G_{d,n}^{+}(x_{1},x_{2})-G_{d,n}^{-}(x_{1},x_{2})\Big]\,.\label{eq:targetfinal}
\end{equation}

As a consistency check with the intermediate description results in Sec.~\ref{sec:review}, let's first compare the ratio between the coefficients of the $G^{+}_{d,n}$ in Eq.~(\ref{eq:targetLL}) and Eq.~(\ref{eq:targetRR}) and the ratio between the coefficients of the $G^{-}_{d,n}$ in Eq.~(\ref{eq:targetLL}) and Eq.~(\ref{eq:targetRR}). For each light field $\hat{O}_{L/R,n}$ we have the two ratios
\begin{equation}
    \frac{\Psi_{n}^{ND}(\mu_{L})^{2}}{\Psi_{n}^{ND\prime}(\mu_{R})^{2}}\,,\quad\frac{\Psi_{n}^{DN\prime}(\mu_{L})^{2}}{\Psi_{n}^{DN}(\mu_{R})^{2}}\,,
\end{equation}
and using (\ref{eq:wronskianresult}) and the fact that $\Psi_{n}^{ND}(\mu_{R})=\Psi_{n}^{DN}(\pi-\mu_{R})=\Psi_{n}^{DN}(\mu_{L})$ we can see that these two ratios are the inverse of the other. This is consistent with the results in Sec.~\ref{sec:review}, i.e. Eq.~(\ref{eq:11}) and Eq.~(\ref{eq:22}), where the corresponding ratios are
\begin{equation}
    \frac{(2\Delta_{-}-d+1)^{2}}{g^{2}}\,,\quad \frac{g^{2}}{(2\Delta_{-}-d+1)^{2}}\,.\label{eq:property2}
\end{equation}
A natural interpretation consistent with this result is that $g_{n}$ is the effective coupling constant for this transparent field, we can see that the squared coupling constant 
is
\begin{equation}
    g_{n}^{2}=(2\Delta_{-}-d+1)^{2}\frac{\Psi_{n}^{ND}(\mu_{L})^{2}}{\Psi_{n}^{ND\prime}(\mu_{R})^{2}}\,,\label{eq:ratio}
\end{equation}
where $\Delta_{-}=\frac{d-1}{2}-\sqrt{\frac{(d-1)^2}{4}+m_{n}^2}$. Furthermore, this result Eq.~(\ref{eq:ratio}) gives the ratio between the coefficient in front of $G^{+}_{d,n}$ in the across AdS$_{d}$ correlator and the coefficient in front of $G^{+}_{d,n}$ in the within right AdS$_{d}$ correlator as
\begin{equation}
    \frac{\Psi_{n}^{ND}(\mu_{L})\Psi_{n}^{ND\prime}(\mu_{R})}{\Psi_{n}^{ND\prime}(\mu_{R})^{2}}= \frac{\Psi_{n}^{ND}(\mu_{L})}{\Psi_{n}^{ND\prime}(\mu_{R})}=\frac{g_{n}}{2\Delta_{-}-d+1}\,,
\end{equation}
which is exactly the same as the same ratio for the results in Sec.~\ref{sec:review}. Second, we can see that the right hand side of the across AdS$_{d}$ Eq.~(\ref{eq:targetfinal}) is consistent with the result in Sec.~\ref{sec:review} for the across AdS$_{d}$ correlator of a free scalar field. Hence in these low-energy operators obey the intermediate description causal structure.  

As a result, we can see that the holographic dictionary gives us results of the intermediate description correlators that are consistent with the intermediate description causal structure at the low-energy sector. This suggests that the Karch-Randall braneworld is indeed a consistent low-energy effective theory model of holography.

\subsection{Holographic Reflection and Transmission}
The intermediate description correlators extracted from the bulk in the last subsection enable us to write down the associated transmission and reflection coefficients following the definitions we gave in Sec.~\ref{sec:tranmrefl}. These coefficients are important physics parameters of the intermediate description. We have the reflection and transmission coefficients associated with each KK mode
\begin{equation}
    R_{n}=\abs{\frac{\Psi_{n}^{ND\prime}(\mu_{R})^{2}+(-1)^{2\Delta_{+}-d+1}\Psi_{n}^{DN}(\mu_{R})^{2}}{\Psi_{n}^{ND\prime}(\mu_{R})^{2}+\Psi_{n}^{DN}(\mu_{R})^{2}}}^{2}\,,
\end{equation}
\begin{equation}
    T_{n}=\abs{\frac{\Psi_{n}^{ND}(\mu_{L})\Psi_{n}^{ND\prime}(\mu_{R})}{\Psi_{n}^{ND\prime}(\mu_{R})^{2}+\Psi_{n}^{DN}(\mu_{R})^{2}}\Big(1-(-1)^{2\Delta_{+}-d+1}\Big)}^{2}\,,
\end{equation}
with $\Delta_{+}=\frac{d-1}{2}+\sqrt{\frac{(d-1)^2}{4}+m_{n}^2}$. As a consistency check, we can see that, using $\Psi_{n}^{ND}(\mu_{L})=\Psi_{n}^{DN}(\mu_{R})$, $R+T=1$ which suggests that a full intermediate description transparent sector is captured by the bulk scalar field. This is consistent with our motivation that the $DN$ and $ND$ modes are sourcing each other. Furthermore, using Eq.~(\ref{eq:ratio}) they can be written as
\begin{equation}
     R_{n}=\abs{\frac{(2\Delta_{+}-d+1)^{2}+(-1)^{2\Delta_{+}-d+1}g_{n}^{2}}{(2\Delta_{+}-d+1)^{2}+g_{n}^{2}}}^{2}\,,
\end{equation}
and 
\begin{equation}
\begin{split}
    T=\abs{g_{n}\frac{(2\Delta_{+}-d+1)}{(2\Delta_{+}-d+1)^{2}+g_{n}^{2}}\Big(1-(-1)^{2\Delta_{+}-d+1}\Big)}^{2}\,,
    \end{split}
\end{equation}
which is consistent with the results Eq.~(\ref{eq:reflection}) and Eq.~(\ref{eq:transmission}).

We note that reflection and transmission with an interface have been widely studied in the literature-- for example \cite{Bachas:2001vj,Quella:2006de,Brunner:2015vva,Bachas:2020yxv,Baig:2024hfc,Gutperle:2024yiz}, and many studies conjectured that it can be useful to understand the island model and can be studied holographically \cite{Bachas:2020yxv,Hollowood:2021lsw,Kruthoff:2021vgv}. The Karch-Randall braneworld is a natural setting to construct and study entanglement islands \cite{Geng:2020qvw,Geng:2020fxl,Geng:2021hlu,Geng:2023zhq} and the above results of the transmission and reflection coefficients establish a concrete framework to study this, and their possible implications for entanglement islands holographically. We leave this for a detailed study for the future.

\subsection{The Unitarity Bound}\label{sec:subtle3}
In our discussion in this section, since we are only interested in the low-energy KK modes, we didn't impose the constraint $\frac{d+1}{2}\geq\Delta_{+}\geq\frac{d-3}{2}$, i.e. $\Delta_{\pm}\geq\frac{d-3}{2}$, as we alluded to in Sec.~\ref{sec:review}. This is called the unitarity bound from the boundary description point of view and in the intermediate description it ensures that the KK modes Eq.~(\ref{eq:convert}) are normalizable under the AdS$_{d}$ Klein-Gordon norm. This bound is generally satisfied by light KK modes but violated by heavy KK modes in the alternative quantization.   For this reason, we will not assume transparency for the high energy modes. In the next section, we will see that only UV modes can be the source of causality violation so with our assumptions, this is never an issue. Without allowing for the unitarity violating modes, the high-energy sector does not give rise to a ``shortcut" so even heavy modes respect causality. However, it is possible that with a cutoff and weak breaking of conformal symmetry in the transparent boundary condition and hence a weaker unitarity constraint, this conclusion about high energy modes could be less robust.

\section{Effects of the Heavy KK Modes}\label{sec:heavy KK modes}
In this section, we study the contribution of heavy KK modes to the two point correlator. There are two questions we want to study for a better understanding of the intermediate description and bulk description causality/locality. The first question is to understand the robustness of the low-energy intermediate description causality we uncovered in Sec.~\ref{sec:Resolution} against interactions between the low-energy fields and the high-energy fields. The second question is to understand the emergence of the bulk causal structure from the intermediate description point of view, which clearly should reflect 5d propagation.  The heavy KK modes play an essential role in these two questions, and so in the interplay between the intermediate description and bulk description causal structures.

\subsection{General Aspects of Heavy KK Modes}\label{sec:largen}
Before addressing the above two questions, let's first outline some useful general features of the heavy KK modes. For the heavy KK modes, $m_{n}^2$ is large and Eq.~(\ref{eq:waveeq}) simplifies to
\begin{equation}
\Psi_{n}''(\mu)+m_{n}^{2}\Psi_{n}(\mu)=0\,,
\end{equation}
which has eigenvalues and normalized wavefunctions
\begin{equation}
     m_{n}=\frac{(n+\frac{1}{2})}{\mu_{R}-\mu_{L}}\pi\,,\quad\Psi_{n}^{ND}(\mu)=\sqrt{\frac{2}{\mu_{R}-\mu_{L}}}\sin m_{n}(\mu_{R}-\mu)\,,\label{eq:KKheavy}
\end{equation}
where $n\in \mathbb{Z}_{+}$ and $n\gg1$. Next, we want to obtain the asymptotic form of $G^{\pm}_{d,n}(x_{1},x_{2})$ in Eq.~(\ref{eq:G}) for these heavy KK modes. This can be done by firstly transforming $G^{+}_{d,n}(x_{1},x_{2})$ to
\begin{equation}
    G^{+}_{d,n}(x_{1},x_{2})=\frac{2^{\frac{d}{2}-1}\Gamma[\Delta_{+}]}{\pi^{\frac{d-1}{2}}\Gamma[\Delta_{+}-\frac{d-3}{2}]}\rho^{\Delta_{+}} \text{}_{2}F_{1}[\Delta_{+},\frac{d-1}{2},\Delta_{+}-\frac{d-3}{2},\rho^{2}]\,,\quad\text{where }\rho=Z-\sqrt{Z^{2}-1}\,,\label{eq:2ptGT}
\end{equation}
for which we first used Eq.(15.8.13) and Eq.(15.8.21) from \cite{NIST:DLMF}. We can secondly use Eq. (15) from \cite{Temme_2003} to get
\begin{equation}
    G^{+}_{d,n}(x_{1},x_{2})\sim\frac{2^{\frac{d}{2}-1}\Gamma[\Delta_{+}]}{\pi^{\frac{d-1}{2}}\Gamma[\Delta_{+}-\frac{d-3}{2}]}\rho^{\Delta_{+}} \frac{1}{(1-\rho^{2})^{\frac{d-1}{2}}}\sim\frac{2^{\frac{d}{2}-1}\Delta_{+}^{\frac{d-3}{2}}}{\pi^{\frac{d-1}{2}}}\rho^{\Delta_{+}} \frac{1}{(1-\rho^{2})^{\frac{d-1}{2}}}\,,\label{eq:Gpasym}
\end{equation}
in the regime where $\Delta_{+}\gg1$. Using the fact that $n\gg1$, we also have
\begin{equation}
    \Delta_{+}\sim  m_{n}\sim\frac{n+\frac{1}{2}}{\mu_{R}-\mu_{L}}\pi\,.\label{eq:asympspec}
\end{equation}

\subsection{Causality and Unitarity in the Intermediate Description}\label{sec:robust}
In Sec.~\ref{sec:Resolution}, we found that the intermediate description correlators are all consistent with the intermediate description causal structure. This makes sense as they are all 4d modes with 4d propagators.

However, we haven't incorporated any interactions. Light fields and heavy fields  interact  so causality for the light fields is not guaranteed in the full theory even if one integrates out the heavy fields as higher derivative operators might be generated in the resulting effective field theory which could be imprints of any potential high-energy causality violation.

 Here we provide a simple model of this phenomenon.  We will see that this type of causality violation is avoided when we restrict our theory to only those high-energy modes consistent with unitarity. Thus, the intermediate description causality in the low-energy regime is robust against possible troubles caused by the bulk geodesic shortcut. We include this here to demonstrate how causality violation would arise potentially in more general theories.  For the intermediate theory, this would only happen if the unitarity conditions as stated are not robust (for reasons such as additional boundary operators as suggested in \cite{Andrade:2011dg,Andrade:2011nh}) or the CFT is somehow broken our model below does suggest a way in principle causality violation could sneak into a low-energy theory.

We now focus on the correlator between the intermediate description operators $\hat{O}_{L,0}(x)$ and $\hat{O}_{R,0}(x)$ corresponding to the lowest KK mode. Let's consider the interaction  between this mode and a sum over heavy modes
\begin{equation}
    \delta \mathcal{L}_{\text{int}}=\lambda\int d^{d}x\sqrt{-g} \hat{\phi}_{L,0}(x)\sum_{n}\hat{\Phi}_{L,n}(x)+\lambda\int d^{d}x\sqrt{-g} \hat{\phi}_{R,0}(x)\sum_{n}\hat{\Phi}_{R,n}(x)\,,\label{eq:modelint}
\end{equation}
where we first take the operators $\hat{O}_{L/R,n}(x)$ to be given as Eq.~(\ref{eq:interoperator}) for both the light and heavy ones.
One can compute the across AdS$_{d}$ connected correlator for the light operator to the leading order in the interaction as
\begin{equation}
      \langle\hat{\phi}_{L,0}(x_{1})\hat{\phi}_{R,0}(x_{2})\rangle=G_{0,0}^{LR}(x_{1},x_{2})+\lambda^{2}\int d^{d}y\sqrt{-g}\int d^{d}w\sqrt{-g}G_{0,0}^{LL}(x_{1},y)\sum_{n}G_{n,0}^{LR}(y,w)G_{0,0}^{RR}(w,x_{2})\,,\label{eq:targetinteraction}
\end{equation}
where $G_{n,0}^{LR}(x_{1},x_{2})$ denotes the free propagator
\begin{equation}
G_{n,0}^{LR}(x_{1},x_{2})=\Psi_{n}^{ND}(\mu_{L})\Psi_{n}^{ND\prime}(\mu_{R})\Big[G_{d,n}^{+}(x_{1},x_{2})-G_{d,n}^{-}(x_{1},x_{2})\Big]\,,\label{eq:GLR}
\end{equation}
so as $G_{n,0}^{LL}(x_1,x_2)$ and $G_{n,0}^{RR}(x_1,x_2)$. From Equ.~(\ref{eq:targetinteraction}) we can see that the effect of the heavy fields to the low-energy correlator is encoded in
\begin{equation}
   f(y,w)\equiv \sum_{n}G_{n,0}^{LR}(y,w)\,.\label{eq:fdef}
\end{equation}
If $f(y,w)$ as a Green function doesn't obey the intermediate description causality then it will induce causality violation in the low-energy sector. Thus, to diagnose whether the  heavy fields can induce causality violation we would study $f(y,w)$ and analyze its branch cut structures. If it has a branch cut terminating on a singularity which corresponds to the case that $y$ and $w$ are spacelike separated in the intermediate description geometry, then causality violation could occur. In fact we will see that there potentially exists such a singularity and it corresponds to the case that $(y,\mu_{L})$ and $(w,\mu_{R})$ are null separated from the bulk perspective. 

The singularity and branch cut structure of each term in Equ.~(\ref{eq:fdef}) has been understood in Sec.~\ref{sec:Resolution} and there is only one branch cut which terminates at a point that $y$ and $w$ are null separated in the intermediate description geometry. A different singularity where a branch cut can end can arise only from an infinite summation. Hence, to see if there is an emergent singularity we can just study  large $n$ contributions. For the sake of convenience, let further define
\begin{equation}
    f^{\pm}(y,w)\equiv\sum_{n}\Psi_{n}^{ND}(\mu_{L})\Psi_{n}^{ND\prime}(\mu_{R})G_{d,n}^{\pm}(y,w)\,,
\end{equation}
that is, we have
\begin{equation}
    f(y,w)=f^{+}(y,w)-f^{-}(y,w)\,.\label{eq:f}
\end{equation}
We can resum the large $n$ contributions to $f^{\pm}(y,w)$ using the results we outlined in Sec.~\ref{sec:largen}. Let's first consider $f^{+}(y,w)$
\begin{equation}
    \begin{split}
        f^{+}(y,w)&\sim\sum_{n}\frac{2n+1}{(\mu_{R}-\mu_{L})^{2}}\frac{(-1)^{n}}{(1-\rho^{2})^{\frac{d-1}{2}}}\frac{\sqrt{2}}{\pi^{\frac{d-3}{2}}}(\frac{\pi}{\mu_{R}-\mu_{L}})^{\frac{d-3}{2}}(2n+1)^{\frac{d-3}{2}}\rho^{\frac{n+\frac{1}{2}}{\mu_{R}-\mu_{L}}\pi}\,,\\&=\frac{1}{(1-\rho^{2})^{\frac{d-1}{2}}}\frac{2^{\frac{d}{2}}}{ (\mu_{R}-\mu_{L})^{\frac{d+1}{2}}}\sum_{n}(-1)^{n}n^{\frac{d-1}{2}}\rho^{\frac{n\pi}{\mu_{R}-\mu_{L}}}\,,\\&=\frac{1}{(1-\rho^{2})^{\frac{d-1}{2}}}\frac{2^{\frac{d}{2}}}{ (\mu_{R}-\mu_{L})^{\frac{d+1}{2}}}\int_{x_{\text{min}}}^{\infty} dx (-1)^{x}x^{\frac{d-1}{2}}e^{x\frac{\pi}{\mu_{R}-\mu_{L}}\log\rho}\,,\label{eq:emergencep}
    \end{split}
\end{equation}
where $\sim$ denotes that we only study the contributions from large $n$ and we have used the asymptotic form of the KK wavefunction and Green functions from Sec.~\ref{sec:largen}. The last step replaces the discrete sum by an integral which can be  justified for large $n$ using Newton's histogram definition of the integral. We will see that the result of the integral in the parameter regime near the singularity is independent of $x_{\text{min}}$ which is a further justification of the above replacement. We are  interested only in the regime such that 
    \begin{equation}
        \log\rho\pm i(\mu_{R}-\mu_{L})\rightarrow0\,.\label{eq:limitrho}
    \end{equation}
The lower bound of the integral $x_{\text{min}}$ in Equ.~(\ref{eq:emergencep}) is a fixed large number which doesn't scale with the parameter $\log\rho$ so we can take $x_{\text{min}}$ as a fixed number while we are taking the limit Equ.~(\ref{eq:limitrho}). Thus the integral Equ.~(\ref{eq:emergencep}) under the limit Equ.~(\ref{eq:limitrho}) becomes 
\begin{equation}
\frac{(\mu_{L}-\mu_{R})^{\frac{d-1}{2}}}{(i(2\mu_{R}-\mu_{2}-\mu_{1})\pm\log\rho)^{\frac{d-1}{2}}}\int_{0}^{\infty} du u^{\frac{d-3}{2}}e^{-u}=\Gamma[\frac{d-1}{2}]\frac{(\mu_{L}-\mu_{R})^{\frac{d-1}{2}}}{(i(2\mu_{R}-\mu_{2}-\mu_{1})\pm\log\rho)^{\frac{d-1}{2}}}\,,
\end{equation}
and $\pm$ depends on if we write $(-1)^{n}$ as $e^{\pm in\pi}$. We also note that the convergence of the above integral is guaranteed by the definition of $\rho$ in terms of the invariant distance $Z$ in Equ.~(\ref{eq:2ptGT}). As a result, we have an emergent singularity for $f^{+}(y,w)$ at Equ.~(\ref{eq:limitrho}) in the form
\begin{equation}
     f^{+}(y,w)
=\frac{\Gamma[\frac{d-1}{2}]}{(1-\rho^{2})^{\frac{d-1}{2}}}\frac{2^{\frac{d}{2}}}{ \mu_{R}-\mu_{L}}\frac{1}{(i(2\mu_{R}-\mu_{2}-\mu_{1})\pm\log\rho)^{\frac{d-1}{2}}}\Big(1+O(\log\rho\pm i(\mu_{R}-\mu_{L}))\Big)\,.\label{eq:singularp}
\end{equation}
The result Eq.~(\ref{eq:singularp}) has a singularity at
\begin{equation}
    \rho=e^{\pm i(\mu_{R}-\mu_{L})}\,.
\end{equation}
In terms of the invariant distance $Z$ these singularities are at
\begin{equation}
 Z=\frac{u_{y}^{2}+u_{w}^{2}-(t_{y}-t_{w})^{2}+i\epsilon\text{ sgn}(t_{y}-t_{w})+(\vec{y}-\vec{w})^{2}}{2u_{y}u_{w}}=\cos(\mu_{R}-\mu_{L})\,,\label{eq:bulksingu1}
\end{equation}
The singularity in Eq.~(\ref{eq:bulksingu1}) can be approached by setting $\epsilon=0$ and it is equivalent to
\begin{equation}
\frac{u_{y}^{2}+u_{w}^{2}-2u_{y}u_{w}\cos(\mu_{R}-\mu_{L})-(t_{y}-t_{w})^{2}+(\vec{y}-\vec{w})^{2}}{2u_{y}u_{2}}=0\,,\label{eq:bulksignal1}
\end{equation}
which is exactly when the two bulk points $(y,\mu_{L})$ and $(w,\mu_{R})$ are connected by a bulk null geodesic. 

Similarly, one can consider $f^{-}(y,w)$ in the same regime Equ.~(\ref{eq:limitrho}) and the result is
\begin{equation}
    \begin{split}
        f^{-}(y,w)&\sim\sum_{n}\frac{2n+1}{(\mu_{R}-\mu_{L})^{2}}\frac{(-1)^{n+\frac{d-3}{2}}}{(1-\rho^{2})^{\frac{d-1}{2}}}\frac{\sqrt{2}}{\pi^{\frac{d-3}{2}}}(\frac{\pi}{\mu_{R}-\mu_{L}})^{\frac{d-3}{2}}(2n+1)^{\frac{d-3}{2}}\rho^{-\frac{n+\frac{1}{2}}{\mu_{R}-\mu_{L}}\pi}\,,\\&=\frac{1}{(1-\rho^{2})^{\frac{d-1}{2}}}\frac{2^{\frac{d}{2}}}{ (\mu_{R}-\mu_{L})^{\frac{d+1}{2}}}\sum_{n}(-1)^{n}n^{\frac{d-1}{2}}\rho^{-\frac{n\pi}{\mu_{R}-\mu_{L}}}\,,\\&=\frac{1}{(1-\rho^{2})^{\frac{d-1}{2}}}\frac{2^{\frac{d}{2}}}{ (\mu_{R}-\mu_{L})^{\frac{d+1}{2}}}\int_{x_{\text{min}}}^{\infty} dx (-1)^{x}x^{\frac{d-1}{2}}e^{-x\frac{\pi}{\mu_{R}-\mu_{L}}\log\rho}\,.\label{eq:emergencem}
    \end{split}
\end{equation}
We note that as opposed to $f^{+}(y,w)$ the above integral is not convergent as one can prove that $Re(\log\rho)<0$. Thus, the integral needs further regularization. One way for the regularization to do analytic continuation from $Re(\log\rho)>0$ to $Re(\log \rho)<0$. Under this analytical continuation, we have
\begin{equation}
    f^{-}(y,w)=\frac{\Gamma[\frac{d-1}{2}]}{(1-\rho^{2})^{\frac{d-1}{2}}}\frac{2^{\frac{d}{2}}}{ \mu_{R}-\mu_{L}}\frac{1}{(i(2\mu_{R}-\mu_{2}-\mu_{1})\pm\log\rho)^{\frac{d-1}{2}}}\Big(1+O(\log\rho\pm i(\mu_{R}-\mu_{L}))\Big)\,,\label{eq:singularm}
\end{equation}
which is exactly the same as $f^{+}(y,w)$. Thus, we can conclude that under the above regularization procedure $f(y,w)$ doesn't have any emergent singularity that could violate the intermediate description causal structure.

Nevertheless, a different quantization or a deformation in Equ.~(\ref{eq:modelint}) could in principle generate a singularity corresponding to the bulk causal structure. However, this is not an issue for the intermediate description when we restrict our formulation to modes consistent with the unitarity bound, as this would remove the trasparent boundary condition for the heavy modes (for which there is not consistent alternative quantization). Heavy fields with the alternative quantization have a large negative conformal weight $\Delta_{-}$ that violates the unitarity bound $\Delta_{-}>\frac{d-3}{2}$. Hence the $\Delta_{-}$ modes cannot be dynamical for the heavy fields. Thus, if we want to properly incorporate the heavy KK modes, the right hand side of Equ.~(\ref{eq:bulkfield}) should be modified for the heavy KK modes such that the $\Delta_{-}$ modes are not dynamical. This can be done by introducing another KK base besides those in Equ.~(\ref{eq:convert})
\begin{equation}
\chi^{b'}_{n,\omega,\vec{k}}(x,\mu)=\sin^{\frac{d-1}{2}}\mu\Psi_{n}^{DN}(\mu) u^{\frac{d-1}{2}}J_{\nu_n}(u\sqrt{\omega^2-\vec{k}^2})e^{-i\omega t+i\vec{k}\dot \vec{x}}\,,\label{eq:convertextra}
\end{equation}
and quantize the bulk field instead as
\begin{equation}
\begin{split}
\hat{\chi}(x,\mu)&=\int\frac{d\vec{k}d\omega}{(2\pi)^{d-1}\sqrt{2\omega}}\sum_{n\in \mathbb{N}_{\text{light}}}\Big[\hat{a}_{n,\omega,\vec{k}}\chi_{n,\omega,\vec{k}}^{a}(x,\mu)+\hat{b}_{n,\omega,\vec{k}}\chi^{b}_{n,\omega,\vec{k}}(x,\mu)+\hat{a}^{\dagger}_{n,\omega,\vec{k}}\chi_{n,\omega,\vec{k}}^{a*}(x,\mu)+\hat{b}^{\dagger}_{n,\omega,\vec{k}}\chi^{b*}_{n,\omega,\vec{k}}(x,\mu)\Big]\,,\\&+\int\frac{d\vec{k}d\omega}{(2\pi)^{d-1}\sqrt{2\omega}}\sum_{n\in \mathbb{N}_{\text{heavy}}}\Big[\hat{a}_{n,\omega,\vec{k}}\chi_{n,\omega,\vec{k}}^{a}(x,\mu)+\hat{b}_{n,\omega,\vec{k}}\chi^{b'}_{n,\omega,\vec{k}}(x,\mu)+\hat{a}^{\dagger}_{n,\omega,\vec{k}}\chi_{n,\omega,\vec{k}}^{a*}(x,\mu)+\hat{b}^{\dagger}_{n,\omega,\vec{k}}\chi^{b'*}_{n,\omega,\vec{k}}(x,\mu)\Big]\,,\label{eq:bulkfieldcareful}
\end{split}
\end{equation}
where $\mathbb{N}_{\text{light}}$ is the set on $n$ such that the unitarity bound is not violated and $\mathbb{N}_{\text{light}}$ is the set on $n$ such that the unitarity bound is violated. We can see that following Equ.~(\ref{eq:interoperator}), the light fields correlators are the same as before and the heavy fields correlators are modified as
\begin{equation}
    \begin{split}
\langle\hat{\Phi}_{L,n}(x_{1})\hat{\Phi}_{L,m}(y_{1})\rangle&=\delta_{mn}\Big[\Psi_{n}^{ND}(\mu_{L})^2G_{d,n}^{+}(x_{1},y_{1})+\Psi_{n}^{DN\prime}(\mu_{L})^{2}G_{d,n}^{+}(x_{1},y_{1})\Big]\,,\\\langle\hat{\Phi}_{R,n}(x_{2})\hat{\Phi}_{R,m}(y_{2})\rangle&=\delta_{mn}\Big[\Psi_{n}^{ND\prime}(\mu_{R})^2G_{d,n}^{+}(x_{2},y_{2})+\Psi_{n}^{DN}(\mu_{R})^{2}G_{d,n}^{+}(x_{2},y_{2})\Big]\,,\\
\langle\hat{\Phi}_{L,n}(x_{1})\hat{\Phi}_{R,m}(x_{2})\rangle&=0\,.\label{eq:targetheavy}
\end{split}
\end{equation}
Hence the interaction between the heavy fields and the light fields in the intermediate description we have described will not induce any violation of the intermediate description causalities and the heavy fields are not transparent any more. Therefore, we can conclude that unitarity protected the low-energy causality of the intermediate description against possible imprints of the bulk geodesic shortcut from the heavy-mass/high-energy sector.

To summarize, all such quantizations will be consistent with causality at low energy. The absence of high-energy contributions depends strongly on the transparent boundary condition, which we have taken to be consistent with unitarity and hence see no violation of intermediate sector causality.

\subsection{The Emergence of  Bulk Locality}
In the above subsection, we saw that  heavy fields are irrelevant for the intermediate description low-energy causality due to our formulation which was designed to maintain unitarity. In this subsection, we will see that the heavy fields are nevertheless essential to reproducing  bulk locality in our formulation.

Bulk locality is probed by the bulk two-point function, which is
\begin{equation}
    \begin{split}
        \frac{\langle\hat{\chi}(x_{1},\mu_{1})\hat{\chi}(x_{2},\mu_{2})\rangle}{\sin^{\frac{d-1}{2}}\mu_{1}\sin^{\frac{d-1}{2}}\mu_{2}}=&\sum_{n\in \mathbb{N}_{\text{light}}}\Big[\Psi_{n}^{ND}(\mu_{1})\Psi_{n}^{ND}(\mu_{2})G_{d,n}^{+}(x_{1},x_{2})+\Psi_{n}^{DN}(\mu_{1})\Psi_{n}^{DN}(\mu_{2})G_{d,n}^{-}(x_{1},x_{2})\Big]\\+&\sum_{n\in \mathbb{N}_{\text{heavy}}}\Big[\Psi_{n}^{ND}(\mu_{1})\Psi_{n}^{ND}(\mu_{2})+\Psi_{n}^{DN}(\mu_{1})\Psi_{n}^{DN}(\mu_{2})\Big]G_{d,n}^{+}(x_{1},x_{2})\,.\label{eq:diccareful}
    \end{split}
\end{equation}
Bulk locality requires that the bulk two-point function is singular when the two bulk points $(x_{1},\mu_{1})$ and $(x_{2},\mu_{2})$ are null separated, i.e. if the AdS$_{d}$ invariant distance obeys $Z(x_{1},x_{2})=\cos(\mu_{1}-\mu_{2})$. As we discussed in Sec.~\ref{sec:Resolution} each term under the sum in Equ.~(\ref{eq:diccareful}) is free from such a singularity. Thus, this singularity can only emerge from the  sum over the contributions from the heavy KK modes. Let's define their contributions as $g_{12}$, i.e.
\begin{equation}
    g_{12}=\sum_{n\in \mathbb{N}_{\text{heavy}}}\Big[\Psi_{n}^{ND}(\mu_{1})\Psi_{n}^{ND}(\mu_{2})+\Psi_{n}^{DN}(\mu_{1})\Psi_{n}^{DN}(\mu_{2})\Big]G_{d,n}^{+}(x_{1},x_{2})\,.
\end{equation}
Using the results from Sec.~\ref{sec:heavy KK modes} we can resum the contributions from the super-heavy KK modes in the same way as in Sec.~\ref{sec:robust}. We have
\begin{equation}
    \begin{split}
       g_{12}&\sim\sum_{n}\frac{2}{(\mu_{R}-\mu_{L})}\sin m_{n}(\mu_{R}-\mu_{1})\sin m_{n}(\mu_{R}-\mu_{2})\frac{2^{\frac{d}{2}-1}}{\pi^{\frac{d-1}{2}}(1-\rho^{2})^{\frac{d-1}{2}}}(\frac{\pi}{\mu_{R}-\mu_{L}})^{\frac{d-3}{2}}(2n+1)^{\frac{d-3}{2}}\rho^{\frac{n+\frac{1}{2}}{\mu_{R}-\mu_{L}}\pi}\,,\\&+\sum_{n}\frac{2}{(\mu_{R}-\mu_{L})}\sin m_{n}(\mu_{L}-\mu_{1})\sin m_{n}(\mu_{L}-\mu_{2})\frac{2^{\frac{d}{2}-1}}{\pi^{\frac{d-1}{2}}(1-\rho^{2})^{\frac{d-1}{2}}}(\frac{\pi}{\mu_{R}-\mu_{L}})^{\frac{d-3}{2}}(2n+1)^{\frac{d-3}{2}}\rho^{\frac{n+\frac{1}{2}}{\mu_{R}-\mu_{L}}\pi}\,,\\&=\frac{1}{(1-\rho^{2})^{\frac{d-1}{2}}}\frac{2^{d-\frac{3}{2}}}{\pi (\mu_{R}-\mu_{L})^{\frac{d-1}{2}}}\sum_{n}n^{\frac{d-3}{2}}\sin\frac{n\pi}{\mu_{R}-\mu_{L}}(\mu_{R}-\mu_{1})\sin\frac{n\pi}{\mu_{R}-\mu_{L}}(\mu_{R}-\mu_{2})\rho^{\frac{n\pi}{\mu_{R}-\mu_{L}}}\\&+\frac{1}{(1-\rho^{2})^{\frac{d-1}{2}}}\frac{2^{d-\frac{3}{2}}}{\pi (\mu_{R}-\mu_{L})^{\frac{d-1}{2}}}\sum_{n}n^{\frac{d-3}{2}}\sin\frac{n\pi}{\mu_{R}-\mu_{L}}(\mu_{L}-\mu_{1})\sin\frac{n\pi}{\mu_{R}-\mu_{L}}(\mu_{L}-\mu_{2})\rho^{\frac{n\pi}{\mu_{R}-\mu_{L}}}\,.
    \end{split}
\end{equation}
Using the fact that 
\begin{equation}
    \sin x=\frac{1}{2i}(e^{ix}-e^{-ix})\,,
\end{equation}
we have
\begin{equation}
\begin{split}
    g_{12}&\sim\frac{-1}{(1-\rho^{2})^{\frac{d-1}{2}}}\frac{2^{d-\frac{7}{2}}}{\pi (\mu_{R}-\mu_{L})^{\frac{d-1}{2}}}\sum_{n}n^{\frac{d-3}{2}}(e^{in\pi\frac{2\mu_{R}-\mu_{1}-\mu_{2}}{\mu_{R}-\mu_{L}}}+e^{-in\pi\frac{2\mu_{R}-\mu_{1}-\mu_{2}}{\mu_{R}-\mu_{L}}}-e^{in\pi\frac{\mu_{2}-\mu_{1}}{\mu_{R}-\mu_{L}}}-e^{-in\pi\frac{\mu_{1}-\mu_{2}}{\mu_{R}-\mu_{L}}}\\&\quad\quad\quad+e^{in\pi\frac{2\mu_{L}-\mu_{1}-\mu_{2}}{\mu_{R}-\mu_{L}}}+e^{-in\pi\frac{2\mu_{L}-\mu_{1}-\mu_{2}}{\mu_{R}-\mu_{L}}}-e^{in\pi\frac{\mu_{2}-\mu_{1}}{\mu_{R}-\mu_{L}}}-e^{-in\pi\frac{\mu_{1}-\mu_{2}}{\mu_{R}-\mu_{L}}})\rho^{\frac{n\pi}{\mu_{R}-\mu_{L}}}\,,\\&=\frac{-1}{(1-\rho^{2})^{\frac{d-1}{2}}}\frac{2^{d-\frac{7}{2}}}{\pi (\mu_{R}-\mu_{L})^{\frac{d-1}{2}}}\int_{x_{\text{min}}}^{\infty} dx x^{\frac{d-3}{2}}(e^{ix\pi\frac{2\mu_{R}-\mu_{1}-\mu_{2}}{\mu_{R}-\mu_{L}}}+e^{-ix\pi\frac{2\mu_{R}-\mu_{1}-\mu_{2}}{\mu_{R}-\mu_{L}}}-e^{ix\pi\frac{\mu_{2}-\mu_{1}}{\mu_{R}-\mu_{L}}}-e^{-ix\pi\frac{\mu_{1}-\mu_{2}}{\mu_{R}-\mu_{L}}}\\&\quad\quad\quad+e^{in\pi\frac{2\mu_{L}-\mu_{1}-\mu_{2}}{\mu_{R}-\mu_{L}}}+e^{-in\pi\frac{2\mu_{L}-\mu_{1}-\mu_{2}}{\mu_{R}-\mu_{L}}}-e^{in\pi\frac{\mu_{2}-\mu_{1}}{\mu_{R}-\mu_{L}}}-e^{-in\pi\frac{\mu_{1}-\mu_{2}}{\mu_{R}-\mu_{L}}}))e^{\frac{x\pi}{\mu_{R}-\mu_{L}}\log\rho}\,.\label{eq:bulkheavyKK}
    \end{split}
\end{equation}
Using the same techniques as in Sec.~\ref{sec:robust}, we have the most singular part of the integral
\begin{equation}
    \begin{split}
g_{12}&\sim-\frac{\Gamma[\frac{d-1}{2}]}{(1-\rho^{2})^{\frac{d-1}{2}}}\frac{2^{d-\frac{7}{2}}}{\pi}\Big[\frac{2}{(i(\mu_{2}-\mu_{1})+\log\rho)^{\frac{d-1}{2}}}+\frac{2}{(i(\mu_{1}-\mu_{2})+\log\rho)^{\frac{d-1}{2}}}\\&\quad\quad\quad\quad-\frac{1}{(i(2\mu_{R}-\mu_{2}-\mu_{1})+\log\rho)^{\frac{d-1}{2}}}-\frac{1}{(-i(2\mu_{R}-\mu_{2}-\mu_{1})+\log\rho)^{\frac{d-1}{2}}}\\&\quad\quad\quad\quad-\frac{1}{(i(2\mu_{L}-\mu_{2}-\mu_{1})+\log\rho)^{\frac{d-1}{2}}}-\frac{1}{(-i(2\mu_{L}-\mu_{2}-\mu_{1})+\log\rho)^{\frac{d-1}{2}}}\Big].
    \end{split}
\end{equation}
From here we can see that we have the emergent singularities at
\begin{equation}
    Z=\cos(\mu_{1}-\mu_{2})\,,\quad\cos(2\mu_{R}-\mu_{1}-\mu_{2})\,,\quad \cos(2\mu_{L}-\mu_{1}-\mu_{2})\,,
\end{equation}
where the first one is exactly the singularity required by bulk locality and the last two are due to the boundary conditions near the branes, i.e. mirror charges. Furthermore, the strength of the singularities is correct as in the small distance limit the bulk two-point function should be in the same form as the two-point function for a massless scalar field in flat space which is algebraic in the distance square with a power $-\frac{d-1}{2}$.

Thus we see that the bulk geometry is  indeed encoded in the heavy fields from the intermediate description perspective. Of course this is not surprising since our modes are the KK modes of the bulk theory so summing their contribution should reproduce the 5d result. We do see nonetheless that the causal structure is associated with high energy modes, and it is only after performing the infinite sum that we see the bulk singularity emerge.

We see that the existence of a singularity corresponding to a shortcut depends
only on the high energy modes, and in particular on the coefficients that
correspond to the particular holographic representation. With different boundary conditions or a different holographic dictionary, it is not clear that these high-energy modes will always cancel. However, unless the heavy modes can obey a transparent boundary condition consistent with unitarity, such effects will not exist and consequently will not leak into the low-energy world. It would be interesting to explore if such a model could be constructed.
We do note that even if such an apparent singularity were to emerge,  the result would be sensitive to the UV cutoff of the 5d theory.

\subsubsection{Emergence of the "bulk" in the Effective Theory}

We have seen that the singularity will not emerge if we have a cutoff on the KK modes (or equivalently a spatial cutoff on 5d space).
The theory is therefore consistent with causality and  commutators of operators that are spatially separated will vanish. 

Nonetheless, we might expect there exists some consequences of the shortcut in the effective theory. Such consequences are at best subtle quantum effects. For example, the across AdS$_{d}$ correlators for different intermediate description fields $\hat{O}_{L/R,n}(x)$ would need to interfere in a particular way that reflects their origin as 5d KK modes. This is clear due to the appearance of the KK wavefunction in these correlators from Equ.~(\ref{eq:GLR}). We expect that if we sum up these two-point functions, the result has a peak when the two points are connected by the bulk geodesic shortcut. This peak should become stronger if we add the contributions from more modes and would eventually become the singularity in Sec.~\ref{sec:robust} we found.\footnote{Though as we discussed, unitarity prevents the existence of such singularity as only finite number of intermediate description fields can have nonzero across-AdS$_{d}$ correlator.} 
For this demonstrating purpose, we will first work in the high-energy regime and ignore the unitarity constraints.

\begin{figure}[h]
\begin{center}
\includegraphics[scale=0.8]{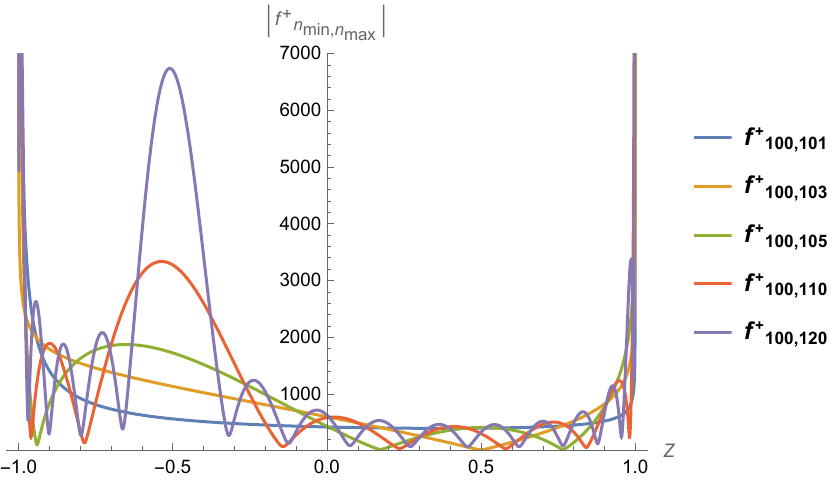}
\caption{The emergence of the ``bulk". $\mu_{L}-\mu_{R}=\frac{2\pi}{3}$}
\label{pic:plots}
\end{center}
\end{figure}

As it follows from Equ.~(\ref{eq:f}), we have the following two contributions to the summation of two-point correlators of the high-energy fields
\begin{equation}
    \begin{split}
        f^{+}(y,w)&\sim\sum_{n}\frac{2n+1}{(\mu_{R}-\mu_{L})^{2}}\frac{(-1)^{n}}{(1-\rho^{2})^{\frac{d-1}{2}}}\frac{\sqrt{2}}{\pi^{\frac{d-3}{2}}}(\frac{\pi}{\mu_{R}-\mu_{L}})^{\frac{d-3}{2}}(2n+1)^{\frac{d-3}{2}}\rho^{\frac{n+\frac{1}{2}}{\mu_{R}-\mu_{L}}\pi}\,,\\&=\frac{1}{(1-\rho^{2})^{\frac{d-1}{2}}}\frac{2^{\frac{d}{2}}}{ (\mu_{R}-\mu_{L})^{\frac{d+1}{2}}}\sum_{n}(-1)^{n}n^{\frac{d-1}{2}}\rho^{\frac{n\pi}{\mu_{R}-\mu_{L}}}\,,
    \end{split}
\end{equation}
and
\begin{equation}
    \begin{split}
        f^{-}(y,w)&\sim\sum_{n}\frac{2n+1}{(\mu_{R}-\mu_{L})^{2}}\frac{(-1)^{n+\frac{d-3}{2}}}{(1-\rho^{2})^{\frac{d-1}{2}}}\frac{\sqrt{2}}{\pi^{\frac{d-3}{2}}}(\frac{\pi}{\mu_{R}-\mu_{L}})^{\frac{d-3}{2}}(2n+1)^{\frac{d-3}{2}}\rho^{-\frac{n+\frac{1}{2}}{\mu_{R}-\mu_{L}}\pi}\,,\\&=\frac{1}{(1-\rho^{2})^{\frac{d-1}{2}}}\frac{2^{\frac{d}{2}}}{ (\mu_{R}-\mu_{L})^{\frac{d+1}{2}}}\sum_{n}(-1)^{n}n^{\frac{d-1}{2}}\rho^{-\frac{n\pi}{\mu_{R}-\mu_{L}}}\,.
    \end{split}
\end{equation}
As an explicit example, let's take $\mu_{R}-\mu_{L}=\frac{2\pi}{3}$ and $d=4$. They are of similar form, and we will focus on $f^{+}(y,w)$. We define
\begin{equation}
    f^{+}_{n_{min},n_{max}}(y,\omega)=\frac{1}{(1-\rho^{2})^{\frac{d-1}{2}}}\frac{2^{\frac{d}{2}}}{ (\mu_{R}-\mu_{L})^{\frac{d+1}{2}}}\sum_{n=n_{min}}^{n=n_{max}}(-1)^{n}n^{\frac{d-1}{2}}\rho^{\frac{n\pi}{\mu_{R}-\mu_{L}}}\,.
\end{equation}
To demonstrate the emergence of a peak when the two points $(y,\mu_{L})$ and $(w,\mu_{R})$ are connected by the bulk geodesic shortcut, i.e. when $Z=\frac{\rho+\rho^{-1}}{2}=\cos(\mu_{R}-\mu_{L})=-0.5$, we will plot the absolute value of $f^{+}_{n_{min},n_{max}}(y,w)$ for $n_{min}=100$ and $n_{max}=101,103,105,110$ and $120$. The results are presented in Fig.~\ref{pic:plots}. As we can see, the more modes we include, i.e. with bigger $n_{max}$, we will have a stronger peak at $Z=\cos(\mu_{R}-\mu_{L})=-0.5$ which is the place where the two brane points are connected by the bulk geodesic shortcut. This peak comes from the interference between different modes and in general its emergence doesn't  rely only on the heavy modes since a sufficient number of light modes also contribute through interference, resulting in such a peak. Here we don't have a good analytic control of the light modes, so we instead used the heavy modes to demonstrate this effect.

However, as we have discussed at the end of Sec.~\ref{sec:robust}, unitarity prevents the heavy modes from participating in the leaky boundary condition and so in our quantization they don't contribute to the summation of the across-AdS$_{d}$ correlators we performed. Only those modes which obey the unitarity condition $\frac{d+1}{2}\geq\Delta_{+}\geq\frac{d-3}{2}$ can obey the transparent boundary condition and hence contribute to the summation. These unitary modes are actually very rare. Thus, we expect the above signal of the bulk to be generally  weak as we wouldn't have enough number of modes that are interfering. To demonstrate this fact, let's consider a particular case where we have the analytic control. Let's consider the case of one brane sitting at $\mu=\frac{\pi}{2}$ with a bulk scalar field having mass $m^{2}=-4$ at $d=4$. The bulk field obeys the normalizable boundary condition near the bulk asymptotic boundary and Neumann boundary condition near the brane. This bulk scalar field obeys the KK wave equation Equ.~(\ref{eq:waveeq}),
whose general solution is given by
\begin{equation}
    \begin{split}
        A&\sin^{\frac{5}{2}-\Delta}(\mu){   }_{2}F_{1}(\frac{5-2\Delta-\sqrt{9+4m_{n}^2}}{4},\frac{5-2\Delta+\sqrt{9+4m_{n}^2}}{4},\frac{1}{2},\cos^{2}\mu)\\&+B\sin^{\frac{5}{2}-\Delta}(\mu)\cos(\mu){  }_{2}F_{1}(\frac{7-2\Delta-\sqrt{9+4m_{n}^2}}{4},\frac{7-2\Delta+\sqrt{9+4m_{n}^2}}{4},\frac{3}{2},\cos^{2}\mu)\,,
    \end{split}
\end{equation}
where we have $\Delta=2+\sqrt{4+m^2}=2$ as in the standard AdS/CFT. The Neumann boundary condition near brane $\Psi'(\frac{\pi}{2})=0$ and the normalizability of the mode tell us that
\begin{equation}
B=0\,,\quad\text{and}\quad m_{n}^2=(2k+\Delta)(2k+\Delta-3)=(2k+2)(2k-1)\,,\text{where}\quad k=0,1,2,3,\cdots\,.\label{eq:Bn}
\end{equation}
Thus, only the $k=0$ mode obeys the unitarity condition $2\geq\Delta_{+}=2\geq0$.  In general with this quantization we expect only at most a few modes participating in the transparent boundary condition. However, we leave open the possibility that alternative quantizations could involve a large (but likely finite) number of modes.

\begin{figure}[h]
\begin{center}
\includegraphics[scale=0.8]{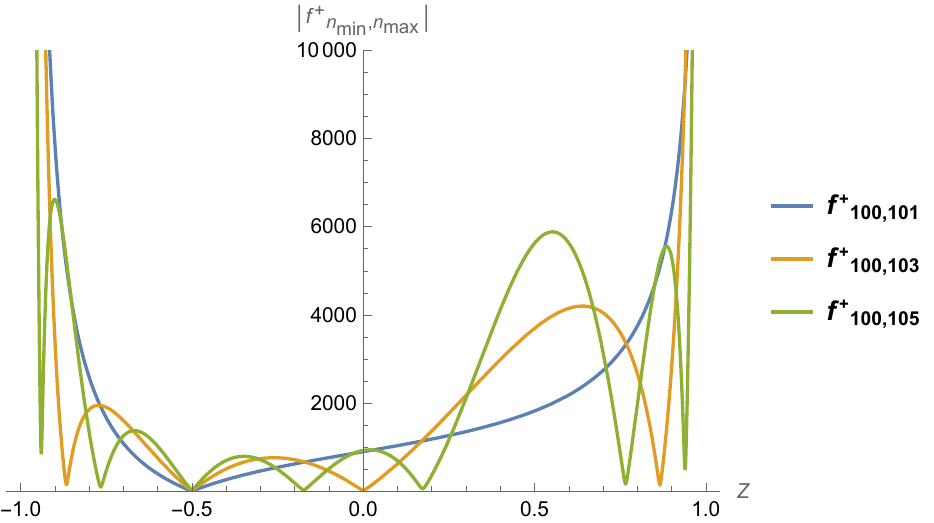}
\caption{The emergence of the ``bulk". $\mu_{L}-\mu_{R}=\frac{\pi}{3}$}
\label{pic:plotssmaller}
\end{center}
\end{figure}

We do expect in general that summing over more modes would lead to enhanced correlators. 
This can be thought of as entanglement deduced from the geometry. 
 We can for example compare the emergent peak  when the branes are closer to each other. We find one needs fewer modes to see its emergence (see Fig.~\ref{pic:plotssmaller}) but in this case the mass gap between nearby modes is bigger.

\section{Other Bulk Geometries and Brane Configurations}\label{sec:other}
So far, we have focused on empty AdS$_{d+1}$ with branes satisfying $\mu_{L}=\pi-\mu_{R}$. We found that the intermediate description low-energy causal structure is obeyed by  intermediate description two-point functions and it is robust against possible interaction with heavy fields when we impose unitarity. These two-point functions are extracted from the bulk following the holographic dictionary we developed. Such holographic dictionaries should extend to more general bulk geometries as long the bulk geometry obeys the following conditions:
\begin{itemize}
    \item The intermediate description is healthy in that the geometries of the Karch-Randall brane don't have closed timelike curves and other acausal structures; 
    \item The brane geometries are asymptotically AdS$_{d}$. Therefore, with two Karch-Randall branes, from the intermediate description point of view, we have two asymptotically AdS$_{d}$ spacetimes that are coupled to each other along their common asymptotic boundary.
    \item The brane configurations in the asymptotic region obey $\mu_{L}=\pi-\mu_{R}$ so that we can have degenerate ND and DN modes in the asymptotic region.
\end{itemize}
These follow because the transparent boundary condition is  sensitive only to the asymptotic geometry near the defect. So as long as the asymptotic geometry is fixed, we can quantize the bulk quantum fields  as above  in the asymptotic region and then extend  into the full bulk using equations of motion. With this prescription, the intermediate description causal structure is preserved in these more general bulk geometries.

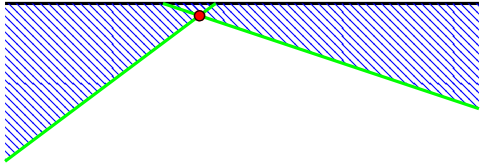
\begin{figure}
\begin{centering}
\begin{tikzpicture}[scale=1.4]
\draw[-,very thick,black!100] (-2,0) to (0,0);
\draw[-,very thick,black!100] (0,0) to (2.5,0);
\draw[pattern=north west lines,pattern color=blue!200,draw=none] (0,0) to (-2,-1.5) to (-2,0) to (0,0);
\draw[pattern=north west lines,pattern color=blue!200,draw=none] (-0.5,0) to (2.5,-1) to (2.5,0) to (-0.5,0);
\draw[-,very thick,color=green!!50] (0,0) to (-2,-1.5);
\draw[-,very thick,color=green!!50] (-0.5,0) to (2.5,-1);
\node at (-0.15,-0.12) {\textcolor{red}{$\bullet$}};
\node at (-0.15,-0.12) {\textcolor{black}{$\circ$}};
\end{tikzpicture}
\caption{A possible configuration with two branes at different ``angles" with the defect conformal symmetry breaking.}
\label{pic:demonbroken}
\end{centering}
\end{figure}

Nevertheless, our formulation doesn't apply when there are two aymmetric branes. In these cases, the ND and DN modes are no longer guaranteed to have the same KK spectrum. This clearly  requires some modification of our prescription.

If we assume that the intermediate description transparent boundary condition has a weakly coupled description as we have considered, i.e. with the conformal symmetry of the boundary description preserved,  two sets of degenerate KK modes are required. This can be achieved by having two bulk scalar fields with equal mass and obeying the same boundary conditions near the branes. The detailed dictionary between the bulk fields and brane operators will be different depending on the boundary conditions. We consider one such possibility in the Appendix.  

The second possibility is that the boundary conformal symmetry is necessarily broken in the asymmetric case. Thus, if the transparent boundary condition still has a weakly coupled description as a double-trace deformation, this deformation wouldn't necessarily be marginal and would correspond in general to nondegenerate modes. This formulation would require a spatial cutoff near the defect in the bulk to reflect the breaking of the asymptotic conformal symmetry. A possible configuration is shown in Fig.~\ref{pic:demonbroken}. Moreover, the holographic dictionary would be difficult to derive from a low-energy perspective as the bulk geometry is now changed. However, since one can imagine deforming a symmetric configuration by a small amount, it is likely there would be a consistent formulation. If so,  the general lessons about the intermediate description causality would still apply.

\section{Conclusions}\label{sec:conclusion}

In this paper, we performed a systematic study of holography in the Karch-Randall braneworld. One motivation was to better understand the apparent causality puzzle in the Karch-Randall braneworld \cite{Geng:2020np}. This puzzle stems from the existence of a bulk geodesic shortcut connecting two points living on different branes in the Karch-Randall braneworld, as it suggests potentially inconsistent causal structures between the intermediate description and the bulk description. We have seen that this bulk geodesic shortcut does not affect the existence of a low-energy causal intermediate description. 

We developed a holographic dictionary for the Karch-Randall braneworld \cite{Geng:2023qwm}  in the special case of symmetric branes with which we could  extract correlators of probe operators in the intermediate description from the bulk description. We further noticed that the low-energy intermediate description causal structure is robust, even against possible interactions between the light and heavy fields. This might be viewed as a toy model to demonstrate the general lesson that the low-energy causal structure in the intermediate description is immune to any form of the bulk geodesic shortcut and that a consistent quantization of the bulk fields with the intermediate description causality can exist. More generally, we expect that any low-energy field in the intermediate description should reflect the intermediate description causal structure. Only in the UV can violations of intermediate description causality potentially emerge. Any such effect would be sensitive to how the intermediate theory is completed in the UV and other constraints like unitarity.

We realized the transparent boundary condition in different ways in each holographic description. In 3d the CFT operators interact. In 4d the brane operators source each other. And in 5d the bulk fields create modes on the two branes.  From a 4d brane perspective the modes are entangled.

However,  there exists the signature of the bulk in the intermediate description. An explicit example we identified are the extraordinary peak of the magnitude in the sum of various correlators. This is a quantum phenomenon due to the interference between the wavefunctions of different KK modes and the bulk would be encoded in this quantum interference. Interestingly, such a signature of the bulk geometry from the intermediate description entanglement is a realization of the proposal in \cite{VanRaamsdonk:2010pw,Maldacena:2013xja}. 

In summary, we conclude that in general a good causal low-energy description, the possibility of enhanced correlators due to the bulk geometry, and a UV-completion-dependent question about possible emergence of violations of causality are interesting features of the intermediate description of the KR braneworld.

Nonetheless, many questions are still unresolved. The holographic dictionary we developed in Sec.~\ref{sec:Resolution} works only when the two branes are symmetric. Clearly more work needs to be done to understand the holographic dictionary for the intermediate picture more generally. We provide a few attempting calculations in the Appendix.~\ref{sec:Appendix}. Some recent work \cite{Uhlemann:2021itz,Uhlemann:2021nhu,Demulder:2022aij,Deddo:2023oxn,Coccia:2021lpp,Anastasi:2025puv} might be useful starting points to find an analyzable UV complete construction of the Karch-Randall braneworld to address the above questions from a top-down perspective.

\section*{Acknowledgements}
We are grateful to Andreas Karch, Joseph Minahan, Rob Myers, Massimo Porrati, Suvrat Raju and Christoph Uhlemann for useful discussions and helpful comments at various stages of this project. We thank Andreas Karch, Massimo Porrati, Suvrat Raju, Mark Riojas and Christoph Uhlemann for comments on an early version of the draft. We also thank Rashmish Mishra, Mark Riojas and Zixia Wei for discussions. HG would like to thank the hospitality from the Aspen Center for Physics where the final stage of this work is performed. Research at the Aspen Center for Physics is supported by the National Science Foundation grant PHY-2210452 and a grant from the Simons Foundation (1161654, Troyer). The work of HG and LR is supported by a grant from the Physics Department at Harvard University. 
\appendix

\section{An Alternative Holographic Dictionary}\label{sec:Appendix}
In this appendix, we develop an alternative holographic dictionary to extract the intermediate description correlators from the bulk. These alternative holographic dictionary could work for generic two-brane configurations with the transparent sector of the intermediate description modeled by free fields. 

\subsection{A Suggestion for an Alternative Dictionary}
The holographic dictionary to extract the intermediate description correlators from the bulk is defined by a proper quantization of the bulk field and a recipe to extract the intermediate description operators from the bulk operators.   In the above holographic dictionary we considered only one bulk field $\chi(x,\mu)$ as a minimal setup. 

Here we propose an alternative dictionary  in which we treat the two coupled AdS$_{d}$'s separately and we model the holographic dual of the operators in the two AdS$_{d}$'s by two separate quantum fields $\chi_{L}(x,\mu)$ and $\chi_{R}(x,\mu)$ in the AdS$_{d+1}$. In resonance with the standard braneworld holography, we can think of the intermediate description operators as purely source operators with dynamics induced from integrating out some hidden matter fields. We assume the induced dynamics of the sources on the two AdS$_{d}$'s satisfy a transparent boundary condition near their common asymptotic boundary. Thus, we only include $NN$-type modes for the bulk fields with the bulk field  quantized such that the result is consistent with transparent boundary condition between these $NN$-type modes. For the above purpose, we will take the two bulk fields $\chi_{L}(x,\mu)$ and $\chi_{R}(x,\mu)$ to have the same mass. In analogy to Eq.~(\ref{eq:convert}), we first provide the new KK basis
\begin{equation}
\begin{split}
\chi^{a}_{n,\omega,\vec{k}}(x,\mu)=\sin^{\frac{d-1}{2}}\mu\Psi_{n}^{NN}(\mu) u^{\frac{d-1}{2}}J_{\nu_n}(u\sqrt{\omega^2-\vec{k}^2})e^{-i\omega t+i\vec{k}\dot \vec{x}}\,,\\\chi^{b}_{n,\omega,\vec{k}}(x,\mu)=\sin^{\frac{d-1}{2}}\mu\Psi_{n}^{NN}(\mu) u^{\frac{d-1}{2}}J_{-\nu_n}(u\sqrt{\omega^2-\vec{k}^2})e^{-i\omega t+i\vec{k}\dot \vec{x}}\,.\label{eq:convertA}
\end{split}
\end{equation}
We also have to implement the unitarity bound $\Delta_{\pm}>\frac{d-3}{2}$. We can quantize the bulk scalar fields as 
the bulk fields are quantized as
\begin{equation}
\begin{split}
\hat{\chi}_{L}(x,\mu)&=\int\frac{d\vec{k}d\omega}{(2\pi)^{d-1}\sqrt{2\omega}}\Big[\sum_{n\in \mathbb{N}_{\text{light}}}\Big(\hat{a}_{n,\omega,\vec{k}}\chi_{n,\omega,\vec{k}}^{a}(x,\mu)+\hat{b}_{n,\omega,\vec{k}}\chi^{b}_{n,\omega,\vec{k}}(x,\mu)+h.c.\Big)\\&\quad\quad\quad\quad\quad\quad\quad\quad\quad\quad\quad\quad\quad\quad\quad\quad+2\sum_{n\in\mathbb{N}_{\text{heavy}}}\Big(\hat{a}_{n,\omega,\vec{k}}\chi^{a}_{n,\omega,\vec{k}}(x,\mu)+h.c.\Big)\Big]\,,\\
\hat{\chi}_{R}(x,\mu)&=\int\frac{d\vec{k}d\omega}{(2\pi)^{d-1}\sqrt{2\omega}}\Big[\sum_{n\in\mathbb{N}_{\text{light}}}\Big(\hat{a}_{n,\omega,\vec{k}}\chi_{n,\omega,\vec{k}}^{a}(x,\mu)-\hat{b}_{n,\omega,\vec{k}}\chi^{b}_{n,\omega,\vec{k}}(x,\mu)+h.c.\Big)\\&\quad\quad\quad\quad\quad\quad\quad\quad\quad\quad\quad\quad\quad\quad\quad\quad+2\sum_{n\in\mathbb{N}_{\text{heavy}}}\Big(\hat{b}_{n,\omega,\vec{k}}\chi^{a}_{n,\omega,\vec{k}}(x,\mu)+h.c.\Big)\Big]\,,\label{eq:bulkfieldAU}
\end{split}
\end{equation}
where we denote the set of $n$'s that obey the unitarity bound $\Delta_{\pm}>\frac{d-3}{2}$ as $\mathbb{N}_{\text{light}}$ and those violate the unitarity bound as $\mathbb{N}_{\text{heavy}}$. The operator dictionary is
\begin{equation}
     \hat{\Phi}_{L,n}(x)=\frac{\hat{\chi}_{L}(x,\mu_{L})}{\Psi_{n}^{NN}(\mu_{L})\sin^{\frac{d-1}{2}}\mu_{L}}\Big|_{n}\,,\quad\hat{\Phi}_{R,n}(x)=\frac{\hat{\chi}_{R}(x,\mu_{R})}{\Psi_{n}^{NN}(\mu_{R})\sin^{\frac{d-1}{2}}\mu_{R}}\Big|_{n}\,.\label{eq:interoperator2}
\end{equation}
and the various intermediate description two-point functions for light operators can be obtained as
\begin{equation}
\begin{split}
        \langle\Phi_{L,n}(x_1)\Phi_{L,m}(y_{1})\rangle&=\delta_{mn}\Big[G_{d,n}^{+}(x_{1},y_{1})+G_{d,n}^{-}(x_{1},y_{1})\Big]\,,\\\langle \Phi_{R,n}(x_2)\Phi_{R,m}(y_{2})\rangle&=\delta_{mn}\Big[G_{d,n}^{+}(x_{2},y_{2})+G_{d,n}^{-}(x_{2},y_{2})\Big]\,,\\
        \langle \Phi_{L,n}(x_1)\Phi_{R,m}(x_{2})\rangle&=\delta_{mn}\Big[G_{d,n}^{+}(x_{1},x_{2})-G_{d,n}^{-}(x_{1},x_{2})\Big]\,,\label{eq:correlatorsALR}
        \end{split}
\end{equation}
where the $i\epsilon$-prescriptions are the same as before. The various correlators for heavy fields are
\begin{equation}
\begin{split}
        \langle\Phi_{L,n}(x_1)\Phi_{L,m}(y_{1})\rangle&=4\delta_{mn}G_{d,n}^{+}(x_{1},y_{1})\,,\\\langle \Phi_{R,n}(x_2)\Phi_{R,m}(y_{2})\rangle&=4\delta_{mn}G_{d,n}^{+}(x_{2},y_{2})\,,\\
        \langle \Phi_{L,n}(x_1)\Phi_{R,m}(x_{2})\rangle&=0\,.
        \end{split}
        \end{equation}
We note that the dictionary Equ.~(\ref{eq:interoperator2}) is designed such that the KK wavefunction is divided out. This ensures that $R+T=1$ for each transparent sector $n$.

The alternative dictionary we developed above can be generalized to the original setup of the Karch-Randall braneworld, i.e. the setup with only one Karch-Randall braneworld. Here we take the brane to be at an angle $0<\mu_{B}\leq\frac{\pi}{2}$ with positive tension and so the leftover asymptotic boundary of the bulk AdS$_{d+1}$ is at $\mu\rightarrow\pi$.

We use the following KK basis
\begin{equation}
\begin{split}
\chi^{a}_{n,\omega,\vec{k}}(x,\mu)=\sin^{\frac{d-1}{2}}\mu\Psi_{n}^{NS}(\mu) u^{\frac{d-1}{2}}J_{\nu_n}(u\sqrt{\omega^2-\vec{k}^2})e^{-i\omega t+i\vec{k}\dot \vec{x}}\,,\\\chi^{b}_{n,\omega,\vec{k}}(x,\mu)=\sin^{\frac{d-1}{2}}\mu\Psi_{n}^{NS}(\mu) u^{\frac{d-1}{2}}J_{-\nu_n}(u\sqrt{\omega^2-\vec{k}^2})e^{-i\omega t+i\vec{k}\dot \vec{x}}\,,\label{eq:convertBCFT}
\end{split}
\end{equation}
where the KK wavefunction $\Psi_{n}^{NS}(\mu)$ obeys the Neumann boundary condition near the brane $\mu=\mu_{B}$ and standard quantization condition near the asymptotic boundary
\begin{equation}
    \Psi^{NS}(\mu)\sim \sin ^{\Delta-\frac{d-1}{2}}\mu \big(1+\phi(\sin^{2}\mu)\big)\,,\quad\text{as }\mu\rightarrow\pi\,,
\end{equation}
with $\Delta=\frac{d}{2}+\sqrt{\frac{d^2}{4}+m^{2}}$ and the bulk scalar fields $\chi_{L}(x,\mu)$ and $\chi_{R}(x\,\mu)$ have the same mass square $m^{2}$. The bulk fields are quantized as
\begin{equation}
\begin{split}
\hat{\chi}_{L}(x,\mu)&=\int\frac{d\vec{k}d\omega}{(2\pi)^{d-1}\sqrt{2\omega}}\Big[\sum_{n\in \mathbb{N}_{\text{light}}}\Big(\hat{a}_{n,\omega,\vec{k}}\chi_{n,\omega,\vec{k}}^{a}(x,\mu)+\hat{b}_{n,\omega,\vec{k}}\chi^{b}_{n,\omega,\vec{k}}(x,\mu)+h.c.\Big)\\&\quad\quad\quad\quad\quad\quad\quad\quad\quad\quad\quad\quad\quad\quad\quad\quad+2\sum_{n\in\mathbb{N}_{\text{heavy}}}\Big(\hat{a}_{n,\omega,\vec{k}}\chi^{a}_{n,\omega,\vec{k}}(x,\mu)+h.c.\Big)\Big]\,,\\
\hat{\chi}_{R}(x,\mu)&=\int\frac{d\vec{k}d\omega}{(2\pi)^{d-1}\sqrt{2\omega}}\Big[\sum_{n\in\mathbb{N}_{\text{light}}}\Big(\hat{a}_{n,\omega,\vec{k}}\chi_{n,\omega,\vec{k}}^{a}(x,\mu)-\hat{b}_{n,\omega,\vec{k}}\chi^{b}_{n,\omega,\vec{k}}(x,\mu)+h.c.\Big)\\&\quad\quad\quad\quad\quad\quad\quad\quad\quad\quad\quad\quad\quad\quad\quad\quad+2\sum_{n\in\mathbb{N}_{\text{heavy}}}\Big(\hat{b}_{n,\omega,\vec{k}}\chi^{a}_{n,\omega,\vec{k}}(x,\mu)+h.c.\Big)\Big]\,,\label{eq:bulkfieldBCFT}
\end{split}
\end{equation}
with the commutation relations between the creation and annihilation operators determined by the bulk canonical quantization condition. The intermediate description operators can be extracted from the bulk operators as
\begin{equation}
     \hat{\Phi}_{B,n}(x)=\frac{\hat{\chi}_{L}(x,\mu_{B})}{\Psi_{n}^{NS}(\mu_{B})\sin^{\frac{d-1}{2}}\mu_{B}}\Big|_{n}\,,\quad\hat{\Phi}_{bdy}(x)=\lim_{\mu\rightarrow\pi}\frac{\hat{\chi}_{R}(x,\mu)}{\sin^{\Delta}\mu}\,,\label{eq:interoperatorBCFT}
\end{equation}
where $\hat{\Phi}_{B,n}(x)$ is the operator that lives on the gravitating AdS$_{d}$ and $\hat{\Phi}_{bdy}(x)$ is the operators living in the nongravitational bath. We emphasize that the geometry of the nongravitational bath is also AdS$_{d}$ which is Weyl equivalent to a half Minkowski space and the operator $\hat{\Phi}_{bdy}(x)$ is a primary operator in the dual BCFT$_{d}$.

If one Weyl transforms the bath to half Minkowski space then the bath operator is
\begin{equation}
    \hat{\Phi}_{bath}(x)=\lim_{\mu\rightarrow\pi}\frac{\hat{\chi}_{R}(x,\mu)}{u^{\Delta}\sin^{\Delta}\mu}\,.\label{eq:interoperatorBath}
\end{equation}
The various intermediate description two-point functions involving light fields can be calculated as
\begin{equation}
        \langle \Phi_{B,n}(x_1)\Phi_{B,m}(y_{1})\rangle=\Big[G_{d,n}^{+}(x_{1},y_{1})+G_{d,n}^{-}(x_{1},y_{1})\Big]\,,\label{eq:correlatorsBB}
\end{equation}

\begin{equation}
\begin{split}
        \langle \Phi_{B,n}(x_1)\Phi_{bath}(x_{2})\rangle=\lim_{\mu\rightarrow\pi}\frac{\Psi_{n}(\mu)}{\sin^{\Delta}\mu}\Big[\frac{G_{d,n}^{+}(x_{1},x_{2})-G_{d,n}^{-}(x_{1},x_{2})}{u_{2}^{\Delta}}\Big]\,,\label{eq:correlatorsBbath}
        \end{split}
\end{equation}
where the $i\epsilon$-prescriptions are the same as before. The correlators involving heavy fields are
\begin{equation}
        \langle \Phi_{B,n}(x_1)\Phi_{B,m}(y_{1})\rangle=4G_{d,n}^{+}(x_{1},y_{1})\,,\label{eq:correlatorsBBheavy}
\end{equation}
\begin{equation}
\begin{split}
         \langle \Phi_{bath}(x_2)\Phi_{bath}(y_{2})\rangle&=\lim_{\mu_{1},\mu_{2}\rightarrow\pi}\frac{\langle \hat{\chi}_{R}(x_{2},\mu_{1})\hat{\chi}_{R}(y_{2},\mu_{2})\rangle}{u_{1}^{\Delta}u_{2}^{\Delta}\sin^{\Delta}\mu_{1}\sin^{\Delta}\mu_{2}}\\&=\sum_{n\in\mathbb{N}_{\text{light}}}\lim_{\mu\rightarrow\pi}\Big(\frac{\Psi_{n}(\mu)}{\sin^{\Delta}\mu}\Big)^{2}\Big[\frac{G_{d,n}^{+}(x_{2},y_{2})+G_{d,n}^{-}(x_{2},y_{2})}{u_{1}^{\Delta}u_{2}^{\Delta}}\Big]\\&+4\sum_{n\in\mathbb{N}_{\text{heavy}}}\lim_{\mu\rightarrow\pi}\Big(\frac{\Psi_{n}(\mu)}{\sin^{\Delta}\mu}\Big)^{2}\frac{G_{d,n}^{+}(x_{2},y_{2})}{u_{1}^{\Delta}u_{2}^{\Delta}}\,.\label{eq:correlatorsbathbath}
         \end{split}
\end{equation}
An interesting implication of the above result can be obtained by noticing that Eq.~(\ref{eq:correlatorsbathbath}) has a nice BCFT$_{d}$ interpretation as the boundary channel expansion of the correlator between two bulk primary operators with conformal weight $\Delta$. The functions $\frac{G_{d,n}^{\pm}(x_{1},y_{1})}{u_{1}^{\Delta}u_{2}^{\Delta}}$ are the boundary conformal blocks from the boundary primary operators with conformal weights $\Delta_{\pm}=\frac{d-1}{2}\pm\sqrt{\frac{(d-1)^{2}}{4}+m^{2}}$. The coefficient $\lim_{\mu\rightarrow\mu}\frac{\Psi_{n}(\mu)}{\sin^{\Delta}\mu}$ is the boundary operator expansion (BOE) coefficient of the bulk primary operator $\hat{\phi}_{bath}(x)$ with conformal weight $\Delta$ into the boundary primary operators with weights $\Delta_{\pm}=\frac{d-1}{2}\pm\sqrt{\frac{(d-1)^{2}}{4}+m^{2}}$.]]

 We can proceed similarly to above if this alternative quantization is valid. We note that to have the transparent sector captured by a bulk primary operator the boundary operator spectrum should have dual weights $\Delta_{\pm}$ and for a bulk primary operator to capture the transparent sector it should have the same BOE coefficients associated with the boundary primary operators with conformal weights $\Delta_{\pm}$. This is a nontrivial implication to the conformal data of the dual BCFT$_{d}$ for the transparent sector to be modeled by weakly coupled AdS$_{d}$ scalar fields in the intermediate description.

\bibliographystyle{JHEP}

\bibliography{main}

\end{document}